# Using invariant manifolds to construct symbolic dynamics for three-dimensional volume-preserving maps*

Bryan Maelfeyt †, Spencer A. Smith † ‡, and Kevin A. Mitchell †

**Abstract.** Topological techniques are powerful tools for characterizing the complexity of many dynamical systems, including the commonly studied area-preserving maps of the plane. However, the extension of many topological techniques to higher dimensions is filled with roadblocks preventing their application. This article shows how to extend the homotopic lobe dynamics (HLD) technique, previously developed for 2D maps, to volume-preserving maps of a three-dimensional phase space. Such maps are physically relevant to particle transport by incompressible fluid flows or by magnetic field lines. Specifically, this manuscript shows how to utilize two-dimensional stable and unstable invariant manifolds, intersecting in a heteroclinic tangle, to construct a symbolic representation of the topological dynamics of the map. This symbolic representation can be used to classify system trajectories and to compute topological entropy. We illustrate the salient ideas through a series of examples with increasing complexity. These examples highlight new features of the HLD technique in 3D. Ultimately, in the final example, our technique detects a difference between the 2D stretching rate of surfaces and the 1D stretching rate of curves, illustrating the truly 3D nature of our approach.

## 1. Introduction.

Many physical systems are modeled by volume-preserving maps of a three-dimensional space into itself. These include particle advection in incompressible fluid flow [2, 36], mixing of granular media in a tumbler [10], and the motion of charged particles along magnetic field lines in a plasma [4]. Beyond their intrinsic value, volume-preserving maps in three dimensions serve as a stepping stone in our understanding of Hamiltonian (i.e. symplectic) maps in two-dimensions to Hamiltonian maps in four dimensions. For these reasons, volume-preserving maps have been an active area of study, especially over the past two decades. Nevertheless, much remains unknown about the structure of volume-preserving maps, especially compared to the detailed understanding available for area-preserving maps of the plane. For recent context, see the reviews of 2D and 3D transport by Aref et al [2] and by Meiss [36].

Much of the research into volume-preserving maps falls into two broad categories: the study of two-dimensional invariant surfaces, e.g. tori, which serve as barriers to global mixing [10, 20, 37, 44, 49, 50, 58], and the study of two-dimensional invariant (stable and unstable) manifolds, which provide a mechanism for mixing around the invariant surfaces [10, 25, 31–34, 50]. The structure of such manifolds is highlighted by Meiss as one of the key questions

*Submitted to the editors December 5, 2016.

**Funding:** This work was supported by the US DOD, ARO grant W911NF-14-1-0359 under subcontract C00045065-4.

†School of Natural Sciences, University of California, Merced, California, 95343 (bmaelfeyt@ucmerced.edu, smiths@mtholyoke.edu, kmitchell@ucmerced.edu)

‡Department of Physics, Mount Holyoke College, South Hadley, Massachussetts, 01075

guiding future work in chaotic transport [36].

> Questions IX (Higher-Dimensional Trellises). What is the structure of trellises in 4D symplectic maps with homoclinic points on two-dimensional stable and unstable manifolds? *What is the structure of trellises in 3D volume-preserving maps...*

The current work addresses this question in 3D.

In addition to studies of invariant manifolds for 3D maps, important studies have addressed the structure of invariant manifolds for 4D symplectic maps, derived from three-degree-of-freedom Hamiltonian systems. Even the definition of finite-volume resonance zones and their associated lobes is not straightforward for 4D maps (and indeed even for 3D maps) and in many cases such zones and lobes do not exist [5, 21, 56, 57]. In other recent developments, Jung and collaborators [16, 17, 27] have explored the topological structure of fractal chaotic scattering functions for three-degree-of-freedom Hamiltonian systems. Other approaches to higher-dimensional symbolic dynamics are based on topological simplexes [28, 29] and higher-dimensional braids [26].

Over the past several decades, an extensive body of research has developed into the topological structure of maps of the interval [38] and (area-preserving) maps of 2D spaces. Regarding the latter, we are particularly concerned here with the structure of homoclinic or heteroclinic tangles, i.e. intersecting networks of stable and unstable manifolds [18, 19, 45, 46]. In prior work, the technique of homotopic lobe dynamics [9, 39–43, 47] was developed to extract a symbolic description of planar dynamics from finite pieces of 1D stable and unstable manifolds. See the work of Collins [11–15] for a related technique. In these approaches "short time" information contained in the intersection structure of the trellis leads to predictions about what the dynamics must do at "long times" arbitrarily far in the future. Said another way, the existence of certain topological structures, in this case homoclinic or heteroclinic intersections, topologically forces the existence of other topological structures, such as additional intersections or periodic orbits of arbitrarily long period. This is similar to the classic "period three implies chaos" result [30, 48] in which the existence of a period-three orbit for a continuous interval map implies the existence of orbits of any period. The symbolic dynamics resulting from the homotopic lobe dynamics technique describes the evolution of curves in the plane, using homotopy theory. However, one can also develop a partitioning of phase space based on this symbolic dynamics [40]. From this partitioning one can assign symbolic itineraries to trajectories, thereby classifying chaotic trajectories of 2D maps.

The objective of this paper is to extend the homotopic lobe dynamics approach of Refs. [9, 39–43, 47] to continuous maps acting on a three-dimensional space. Our approach is applicable to volume-preserving maps, though volume preservation is not a requirement. (Other topological techniques have been developed to study fractal *attractors* embedded in three dimensions [22].) The input to the technique is (finite-area) pieces of intersecting two-dimensional stable and unstable manifolds attached to hyperbolic fixed points, what we call a *trellis*. The output of the technique is a symbolic dynamics that encodes the topological structure of the map. The mathematical framework for deriving the symbolic dynamics is homotopy theory. To achieve a homotopically nontrivial space, we punch ring-shaped holes in the 3D phase space, adjacent to certain 1D heteroclinic intersections between the stable and unstable manifolds. These intersections are specially selected to topologically force the dynamics encoded within the manifolds. With these ring-shaped holes punched, the homotopy classes describe the manner in which two-dimensional surfaces wrap around the rings. Though it might seem natural to construct our analysis around the second homotopy group $\Pi_2$, which is based on the embedding of spheres within the punctured 3D space, the group $\Pi_2$ is not topologically rich enough for our purposes. Instead, we use homotopy classes based on multiply punctured disks embedded in the punctured 3D space, i.e. on surfaces with an arbitrary (nonzero) number of boundary circles. Though these homotopy classes do not have a well-defined group structure, they do have a well-defined structure of concatenation with one another, which can be represented in a graph-theoretical manner. The symbolic equations are represented by mapping an elemental graph forward to a concatenation of elemental graphs. These equations describe how one elemental homotopy class (what we later call a *bridge class*) is iterated forward to a concatenation of elemental classes. The use of such homotopy classes is, to our knowledge, entirely new, and we have introduced new machinery (e.g. the primary and secondary divisions) and notation (e.g. the graph structures in Fig. 28) to analyze them.

The information in the graph-based dynamical equations can be simplified into a transition matrix, as is common for standard symbolic dynamics techniques, i.e. techniques based on phase space partitioning. From the largest eigenvalue of this matrix, one obtains an entropy for the symbolic dynamics, which is a lower bound on the topological entropy of the original map. It is, in essence, the topological entropy inherent to the trellis used to extract the symbolic dynamics.

Once the symbolic dynamics for the evolution of 2D surfaces has been derived, these equations can be reduced to yield equations for the evolution of 1D curves, based on the fundamental group $\Pi_1$ of the punctured 3D space. This leads to two sets of symbolic dynamics, one for the evolution of 2D surfaces and one for the evolution of 1D curves. Each of these has an associated entropy, what we call the 1D and 2D stretching rates. These stretching rates need not be equal; 2D surfaces can stretch more than 1D curves.

As they evolve, the topology of the 2D stable and unstable manifolds can become very intricate. So, for pedagogical reasons, we have opted in this manuscript to explain the 3D homotopic lobe dynamics technique via a series of explicit examples. These examples are chosen to illustrate the basic concepts and results, and a motivated reader will be able to apply these same techniques to a much wider set of trellises.

This article begins with a discussion of 3D maps with a symmetry axis (Sect. 2), which reduce to 2D maps. This provides an opportunity to review the HLD technique in 2D. We apply this 2D technique to the standard horseshoe topology (Sect. 2.1) and a more complicated horseshoe "with overshoot" (Sect. 2.2). We explain the latter example in a manner that sets the stage for the 3D analysis to come. Sect. 2.3 discusses how the 2D analysis extends to 3D under rotational symmetry. Section 3 presents the analysis of truly 3D maps, without symmetry. Here we focus on three examples of varying levels of complexity. We work through Example 3 (Sect. 3.1) in considerable detail, in order to fully explain the 3D technique. This is the example the reader should focus on to truly understand the technique. Example 4 (Sect. 3.2) is a simpler trellis than Example 3 and illustrates how tangles in 3D can have lower entropy than a corresponding tangle in 2D. Example 5 (Sect. 3.3) is the culmination of this work. It explicitly demonstrates that the 2D stretching rate can be strictly larger than the 1D stretching rate, which is not true for Examples 1-4. In this sense, Example 5 exhibits truly 3D behavior. The reduction of the homotopic dynamics of 2D surfaces to 1D curves is carried out in Sect. 4. Concluding remarks are in Sect. 5.

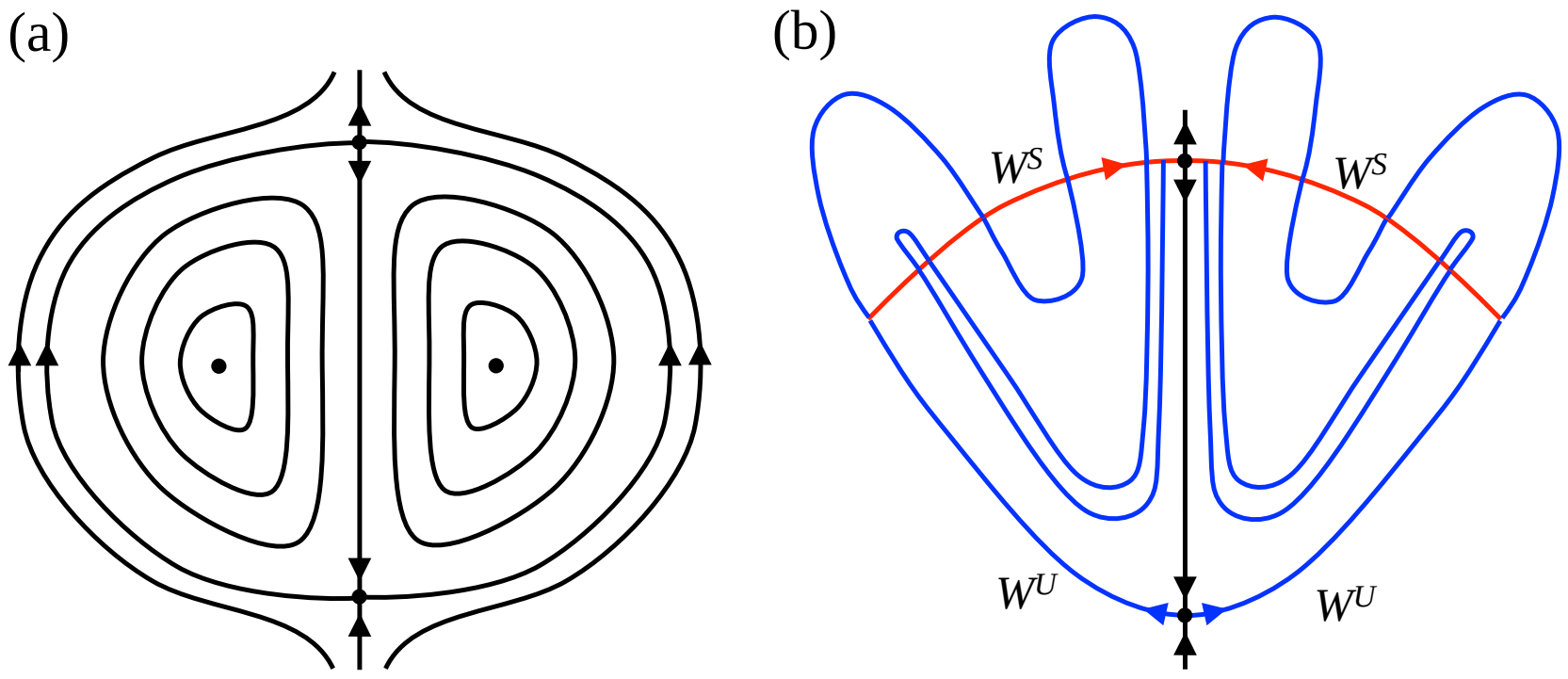

**Figure 1.** *Cross-sections of the 3D phase space for two volume preserving maps with the topology of a spherical ring vortex. Both cases are rotationally invariant about the vertical axis. a) The "integrable" case corresponding to a time-independent flow. The vortex is foliated into invariant tori and a separatrix connects the upper hyperbolic fixed point to the lower hyperbolic fixed point. b) The chaotic case corresponding to a time-periodic flow. The separatrix has broken up into separate stable and unstable manifolds forming a heteroclinic tangle.*

**2. Maps with an axis of symmetry.** We first consider a volume-preserving map with an axis of rotational symmetry. For a concrete physical model, we consider the flow topology of a spherical ring vortex (e.g. Hill's vortex [3, 24]), which is of fundamental interest in 3D fluid mixing studies [35]. Figure 1a shows a cross-sectional sketch of the streamlines in the case of a steady (i.e. time-independent) flow. Note that this phase portrait is a "cartoon" and not based on a specific numerical model; this will be true of all phase portraits in this paper. The fluid flow is rotationally invariant about the $z$ axis, with no component of velocity in the azimuthal direction. Thus, each vertical plane passing through the $z$ axis is invariant under the flow, with identical dynamics to every other such vertical plane. The 3D phase space is foliated into 2D invariant surfaces (invariant curves in the cross-section), with a separatrix connecting a hyperbolic fixed point at the top to a hyperbolic fixed point at the bottom. The top fixed point has two stable directions (horizontal) and one unstable direction (vertical), whereas the bottom fixed point has two unstable directions (horizontal) and one stable direction (vertical). The separatrix separates bound trajectories (living on invariant tori) from unbound trajectories.

Suppose now that the flow is not steady, but *time-periodic*, and that the time-periodic flow retains the rotational symmetry. Consider the Poincaré map $M$ obtained by integrating a given point $\mathbf{x}$ forward under the flow for one period. The map $M$ will then typically exhibit chaos with a much richer phase space structure than the steady case. See Fig. 1b for the sketch of an example phase portrait. For a small enough time-periodic perturbation, the two fixed points persist, but the separatrix in Fig. 1a breaks up into two distinct two-dimensional surfaces, a stable manifold $W^S$ attached to the upper fixed point and an unstable manifold $W^U$ attached to the lower fixed point. These manifolds do not self-intersect, but do intersect one another, forming a complicated *heteroclinic tangle*. Due to the rotational symmetry, this tangle can be analyzed within a cross-sectional plane. We shall carry out this 2D analysis using the homotopic lobe dynamics technique developed previously [9, 39–43, 47]. The present description of the 2D HLD analysis differs in some significant ways from earlier references; the purpose here is to set the stage for the 3D asymmetric case to come in Sect. 3.

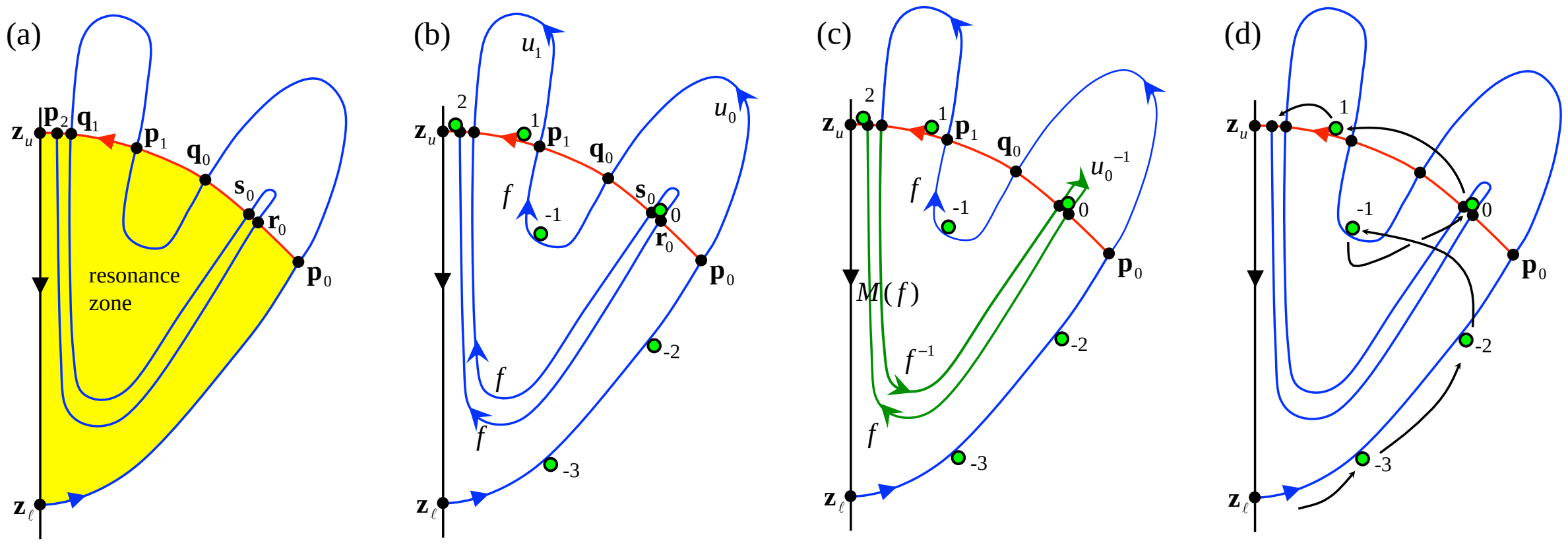

**Figure 2.** *The right half of the tangle from Fig. 1b, used in Example 1. a) The inner region, or resonance zone, defined by the pip* $\mathbf{p}_0$ *is shaded yellow. b) The green-filled dots form a trajectory of holes punched in the plane. c) The green curve, a representative of* $M(f)$*, has a homotopy class that decomposes into the product* $f^{-1}u_0^{-1}f$*. d) The holes can be viewed as rods advancing forward in time according to the black arrows. The arrow from -2 to -1 is shown passing over the arrow from -1 to 0, indicating that the rod from -2 to -1 passes through the intersection point before the rod from -1 to 0.*

**2.1. Example 1: a complete horseshoe.** Due to symmetry, we focus on just the right side of Fig. 1b, which is shown in Fig. 2a with intersection points labelled. The point $\mathbf{p}_0$ is a *primary intersection point (pip)*[18, 19], meaning that $W^S[\mathbf{z}_u, \mathbf{p}_0]$ intersects $W^U[\mathbf{z}_\ell, \mathbf{p}_0]$ only at $\mathbf{p}_0$. Here, the square bracket notation $W^S[\mathbf{x}, \mathbf{y}]$ refers to the (closed) interval of the curve $W^S$ between $\mathbf{x}, \mathbf{y} \in W^S$. Similar notation applies to $W^U$. The two segments $W^S[\mathbf{z}_u, \mathbf{p}_0]$ and $W^U[\mathbf{z}_\ell, \mathbf{p}_0]$, together with the invariant line segment connecting $\mathbf{z}_\ell$ to $\mathbf{z}_u$, bound a *resonance zone* (shaded yellow) forming the interior of the vortex. We next focus on subsequent intersections of the unstable manifold $W^U$ with the fixed interval $W^S[\mathbf{z}_u, \mathbf{p}_0]$ of the stable manifold. The pips $\mathbf{q}_0$ and $\mathbf{p}_1 = M(\mathbf{p}_0)$ define the intervals $W^U[\mathbf{p}_0, \mathbf{q}_0]$ and $W^U[\mathbf{q}_0, \mathbf{p}_1]$, which bound regions called *lobes* that are exterior and interior to the vortex, respectively. Iterates of the exterior lobe are taken to remain outside the vortex (no recapture), whereas iterates of the interior lobe are stretched and folded inside the resonance zone of the vortex, necessarily developing additional intersections with the stable segment $W^S[\mathbf{z}_u, \mathbf{p}_0]$. In the present example, we assume this happens immediately upon the first iterate $W^U[\mathbf{q}_1, \mathbf{p}_2]$, forming a pair of secondary intersections $\mathbf{s}_0$ and $\mathbf{r}_0$. This tangle thus forms a complete horseshoe, with well-known binary symbolic dynamics having topological entropy $\ln 2$.

Given the finite-length pieces of the stable and unstable manifolds in Fig. 2a, what we call the *trellis*, the homotopic lobe dynamics technique determines the minimal topological complexity forced in subsequent iterates of the unstable manifold. For the sake of simplicity and pedagogy, we only sketch how the technique is applied here, leaving a more detailed example to Sect. 2.2.

We first punch a bi-infinite sequence of holes in the plane, shown in Fig. 2b as green-filled circles. Hole 0 is placed within the domain bounded by $W^U[\mathbf{s}_0, \mathbf{r}_0]$ and $W^S[\mathbf{s}_0, \mathbf{r}_0]$, with the other holes being forward or backward iterates of the hole 0. The sequence of holes converge upon $\mathbf{z}_u$ in the forward time direction and $\mathbf{z}_\ell$ in the backward time direction.

We introduce the term *bridge* for a piece of the unstable manifold that begins and ends on $W^S[\mathbf{z}_u, \mathbf{p}_0]$ but does not otherwise intersect $W^S[\mathbf{z}_u, \mathbf{p}_0]$. Bridges are thus the pieces obtained when $W^U$ is cut along $W^S[\mathbf{z}_u, \mathbf{p}_0]$. The bridges are grouped into *bridge classes* based on how they wind around the holes. For example, a bridge of class $f$ winds around hole -1 in the clockwise direction (Fig. 2b). The orientation of the bridge class is denoted with a barbed arrow in the figure. In contrast, the directions along the stable and unstable manifolds defined by the dynamics are denoted by unbarbed, triangular arrows. Note that the bridge-class orientation need not be the same as the dynamical direction. There are a total of three bridges of class $f$ in Fig. 2b. In addition, there are two bridges of class $u_0$, which wind counterclockwise around hole 0, and one bridge of class $u_1$, which winds counterclockwise around hole 1.

The dynamics of the map $M$ naturally generates a dynamics on bridge classes. For example, the bridge $W^U[\mathbf{q}_0, \mathbf{p}_1]$ of class $f$ maps forward to the green curve in Fig. 2c. This curve can in turn be decomposed into three bridges of class $f^{-1}$, $u_0^{-1}$, and $f$. The inverse denotes the reverse orientation of a bridge-class. Expressing concatenation of bridges as a product of bridge classes, we thus have

$$M(f) = f^{-1} {u_0}^{-1} f. \tag{1}$$

Similarly, $u_0$ maps forward to $u_1$. In reality, there is an entire sequence of bridge classes $u_n$, $n \geq 0$, satisfying

$$M(u_n) = u_{n+1}, \quad n \geq 0. \tag{2}$$

Because the $u_n$ classes always map forward to a single class, for an arbitrary number of iterates, they are called *inert* classes, A class like $f$, however, is called an *active* class because it generates multiple classes when iterated forward. Because $f$ is the only active class, and it generates two copies of itself (ignoring inverses), the topological entropy of the symbolic dynamics is

$$h_{\text{top}} = \ln 2, \tag{3}$$

as is well known for a complete horseshoe.

One way to visualize the stretching and folding of the unstable manifold is to consider each hole as a rod placed into the 2D phase space, interpreted as an incompressible fluid. Then, over the course of one period of the fluid flow, each rod moves continuously forward with the fluid, eventually advancing to the position of its iterate under the map. The collection of all the rods moving forward through one period defines a topological braid, with an infinite number of strands. The manner in which the rods braid around one another determines how the phase space is stirred and the nature of the chaotic dynamics. (For prior work on the braiding approach, see Refs. [1, 6–8, 23, 47, 51–55] .) Intuitively, the topology of the trellis should determine the topology of the braid and vice versa. For the trellis in Fig. 2, the braiding of the rods is shown in Fig. 2d. The topology of the unstable manifold can be generated by the action of this stirring; imagine a material line coinciding with a short initial segment of the unstable manifold touching both the lower fixed point $\mathbf{z}_\ell$ and a hole $-n$ arbitrarily close to $\mathbf{z}_\ell$. As time evolves, this hole moves away from $\mathbf{z}_\ell$ advancing to each successive hole after each period, stretching the material line into longer and longer segments of the unstable manifold. Up to hole -2, this line has no folds. However, as the end of the material line advances from hole -2 to -1, it acquires its first fold (the curve $W^U[\mathbf{p}_0, \mathbf{q}_0]$), due to the simultaneous motion of rod -1 up to the position of rod 0. Note that the arrow from -2 to -1 is shown passing *over* the arrow from -1 to 0, indicating that the rod from -2 to -1 passes through the intersection point before the rod from -1 to 0. This ordering is critical for describing the correct braiding of the trajectories. This nontrivial braiding of rod -2 to -1 to 0 is responsible for all of the subsequent folding of the manifold. This stirring picture is a physical way of describing the mechanism behind the chaos and for explaining why the holes are so important to the process. It is introduced here, and again in Sect. 2.3 in the 3D case, for its intuitive insight. However, the braiding picture will not be used explicitly in the construction of the symbolic dynamics.

### 2.2. Example 2: a horseshoe with overshoot.

**2.2.1. Trellis.** Figure 3a shows the sketch of a more complicated trellis. As before, this is a cartoon of a possible trellis, rather than the result of an explicit numerical computation. We shall approach the analysis of this trellis more systematically than Example 1, in order to lay the foundation for the truly 3D analysis of Sect. 3. The interior lobe $W^U[\mathbf{q}_0, \mathbf{p}_1]$ (blue) maps forward to $W^U[\mathbf{q}_1, \mathbf{p}_2]$ (orange), which is stretched more than in Example 1 (Fig. 2a), so that the tip of the lobe folds back and reenters the resonance zone, creating additional secondary intersections $\mathbf{t}_0$ and $\mathbf{x}_0$. That is, the tip of $W^U[\mathbf{q}_1, \mathbf{p}_2]$ has "overshot" the external lobe, yielding more stretching than the complete horseshoe in Example 1. Though this example was analyzed previously using the HLD approach [39, 40], the present analysis differs markedly via the introduction of the primary and secondary divisions.

**2.2.2. Escape-time plots and pseudoneighbors.** We next need to punch holes in the plane. Whereas we simply stated where to punch holes in Example 1, here, we proceed more systematically. We first identify a pair of heteroclinic intersections, called *pseudoneighbors*. The pseudoneighbors are in some sense the most important intersections, as they topologically force all the other intersections to exist. We identify pseudoneighbors through the forward and backward *escape-time plots*. These plots are defined in terms of the unstable and stable *fundamental* segments $W^U[\mathbf{p}_{-1}, \mathbf{p}_0]$ and $W^S[\mathbf{p}_0, \mathbf{p}_1]$ shown as bold segments in Fig. 4a. A fundamental segment of a 1D invariant manifold is any interval bounded by a point $\mathbf{x}$ and its iterate $M(\mathbf{x})$ [19].

The forward escape-time plot records the number of forward iterates (the escape time) required for any point on the unstable fundamental segment $W^U[\mathbf{p}_{-1}, \mathbf{p}_0]$ to map out of the resonance zone. See Fig. 4c. Up to a given maximum iterate, the escape-time plot consists of *escape segments*, over which the escape time is constant, separated by gaps. The endpoints of an escape segment are heteroclinic intersection points. These are labelled in Fig. 4c, and can be read off of $W^U[\mathbf{p}_{-1}, \mathbf{p}_0]$ in Fig. 3b. In this figure, the (red) stable manifold is extended backward three iterates to $\mathbf{p}_{-3}$. Preiterates of the heteroclinic intersections in Fig. 3a land inside the resonance zone before eventually landing on the unstable fundamental segment. "Escape" lobes are designated by color; the blue lobe escapes in one iterate, the red in two, and the green in three. The intersections of the "escape" lobes in Fig. 3b with the fundamental segment $W^U[\mathbf{p}_{-1}, \mathbf{p}_0]$ determine the escape segments in the forward escape-time plot. At higher iterates, escape segments occur in the gaps between segments in Fig. 4c, forming a fractal structure. The number of escape segments ultimately grows exponentially in the escape time, where the growth rate is the topological entropy. In a similar manner, the backward escape-time plot records the number of backward iterates required for any point on the stable fundamental segment to map out of the resonance zone. See Fig. 4b. The ordering of points in Fig. 4b is just the ordering of points in Fig. 3a along the fundamental stable segment $W^S[\mathbf{p}_0, \mathbf{p}_1]$. The endpoints of escape segments in the backward escape-time plot are iterates of the endpoints in the forward escape-time plot, though their ordering is permuted. If $\mathbf{x}_{-n}$ bounds a segment that escapes after $n$ iterates in the forward escape-time plot, then $\mathbf{x}_0$ bounds a segment that escapes after $-n$ iterates in the backward escape-time plot. Note that there must be the same number of segments at iterate $n$ in forward time as there are at iterate $-n$ in backward time.

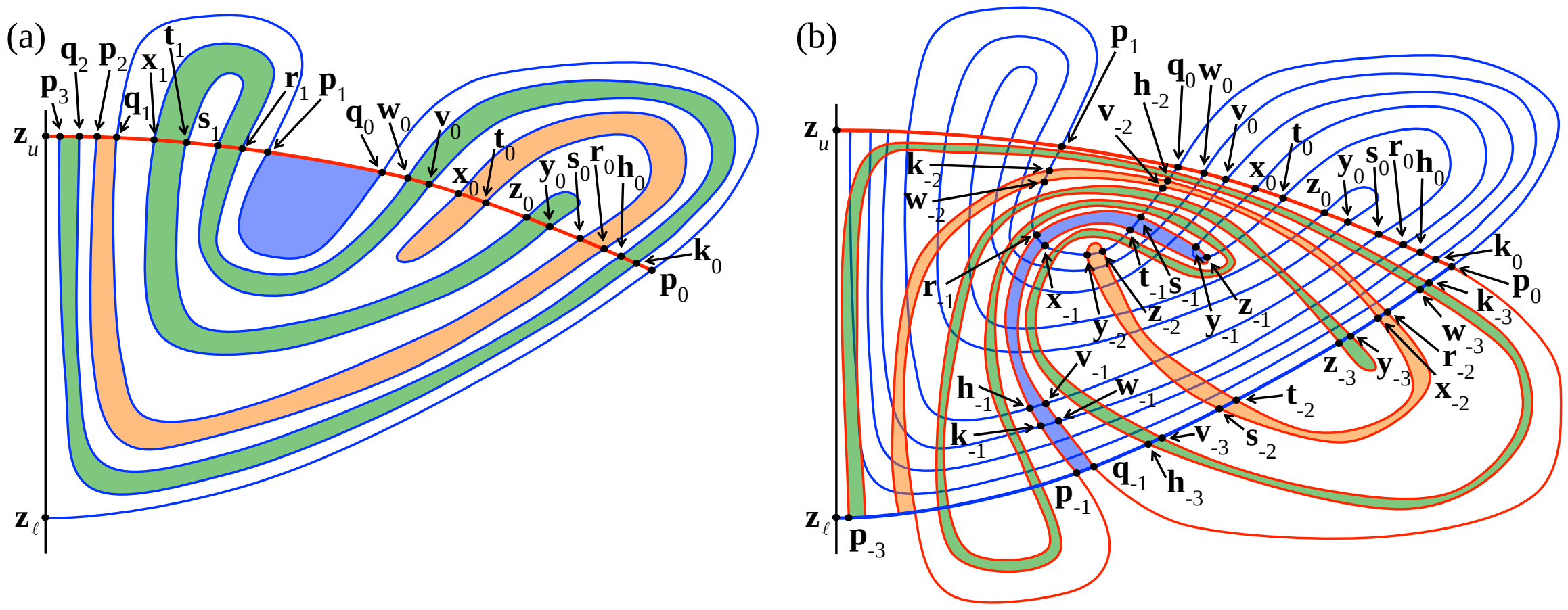

**Figure 3.** *a) The trellis for Example 2. The width has been stretched (relative to Fig. 1b) to better visualize the structure. Successive "capture" lobes are colored blue, orange, and green. b) The stable manifold is continued backward three iterates. Successive "escape" lobes are colored blue, orange, and green.*

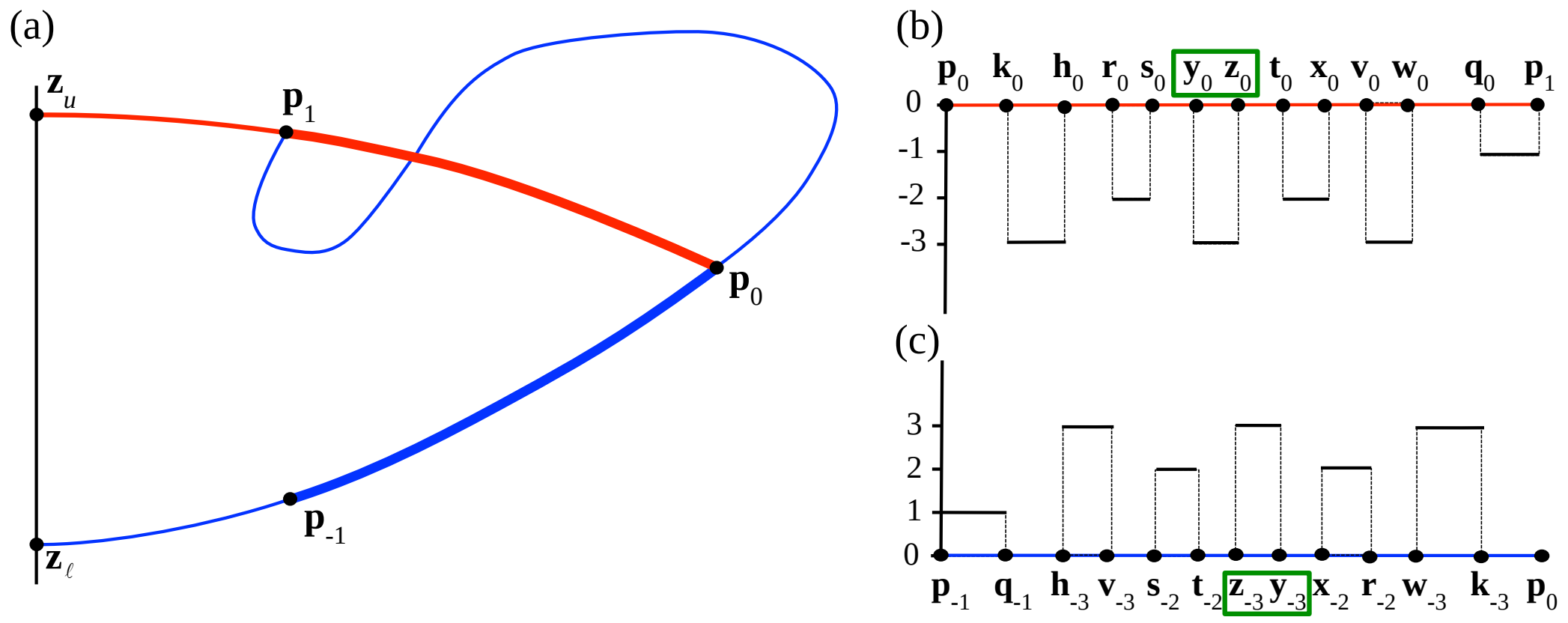

**Figure 4.** *The escape-time plots for Example 2. a) The escape-time plots are defined over the stable and unstable fundamental segments, shown in bold red and blue, respectively. b) The backward escape-time plot. c) The forward escape-time plot. Green boxes enclose the pseudoneighbors.*

If two intersections are adjacent in the forward escape-time plot, i.e. they are endpoints of the same escape segment, and some iterate of these intersections are also adjacent in the backward escape-time plot, then these two intersections form a pseudoneighbor pair; iterates of pseudoneighbors are also considered pseudoneighbors, so we actually have a pair of pseudoneighbor trajectories. In Fig. 4, $\mathbf{y}_n$ and $\mathbf{z}_n$ are pseudoneighbors. In Fig. 5a, the pseudoneighbors $\mathbf{y}_0$ and $\mathbf{z}_0$ define a region bounded by the segments $W^U[\mathbf{y}_0, \mathbf{z}_0]$ and $W^S[\mathbf{y}_0, \mathbf{z}_0]$. If one were to perturb the map $M$ in such a way as to remove some of the heteroclinic intersections from the trellis in Fig. 5a, the pseudoneighbors $\mathbf{y}_0$ and $\mathbf{z}_0$ would have to be removed first in a tangent bifurcation, with the area of the region bounded by $W^U[\mathbf{y}_0, \mathbf{z}_0]$ and $W^S[\mathbf{y}_0, \mathbf{z}_0]$ shrinking to zero. That is, so long as $\mathbf{y}_0$ and $\mathbf{z}_0$ exist, all of the other heteroclinic intersections in Fig. 5a must also exist. This is ultimately why the pseudoneighbors are important. They guarantee, or force, the topological structure of the entire trellis.

**2.2.3. Holes.** We punch a hole inside the region bounded by $W^U[\mathbf{y}_0, \mathbf{z}_0]$ and $W^S[\mathbf{y}_0, \mathbf{z}_0]$, as shown by the green dot in Fig. 5a. Technically, we should punch the hole "infinitesimally" close to one or the other point $\mathbf{y}_0$ or $\mathbf{z}_0$; it does not matter which. This hole is then mapped backward and forward into the bi-infinite sequence of green holes shown in Fig. 5a. If one were to perturb $M$, one could not remove any trellis intersections from Fig. 5a without $W^S$ or $W^U$ passing through one of the holes, which is not allowed. That is, $M$ defined on the punctured plane retains a trellis with the topological structure of Fig. 5a even under continuous distortions.

**2.2.4. Primary division and bridge classes.** As in Example 1 (Sect. 2.1), we define the (homotopy) class of a bridge based on how it winds around the holes. Figure 5b shows three inner bridge classes $a_1$, $a_2$, and $f$, and two outer classes $u_0$ and $u_1$. As in Example 1, there is an infinite sequence of inert bridge classes $u_n$, $n \geq 0$, satisfying

$$M(u_n) = u_{n+1}. \tag{4}$$

We shall call the first class $u_0$ in this sequence a *primary* inert class, and any bridge in this class a *primary* inert bridge.

The reason the inert bridge classes are "uninteresting" is because once a hole lands on $W^S[\mathbf{z}_u, \mathbf{p}_0]$, it simply marches along $W^S[\mathbf{z}_u, \mathbf{p}_0]$ toward the fixed point $\mathbf{z}_u$, generating no new stretching and folding. As with the bridge classes, we call such a hole an *inert* hole. Henceforth, we ignore all inert holes, except the primary ones, i.e. the first such holes to land on $W^S[\mathbf{z}_u, \mathbf{p}_0]$. This simplifies our topological analysis, allowing us to ignore all inert classes, except the primary one.

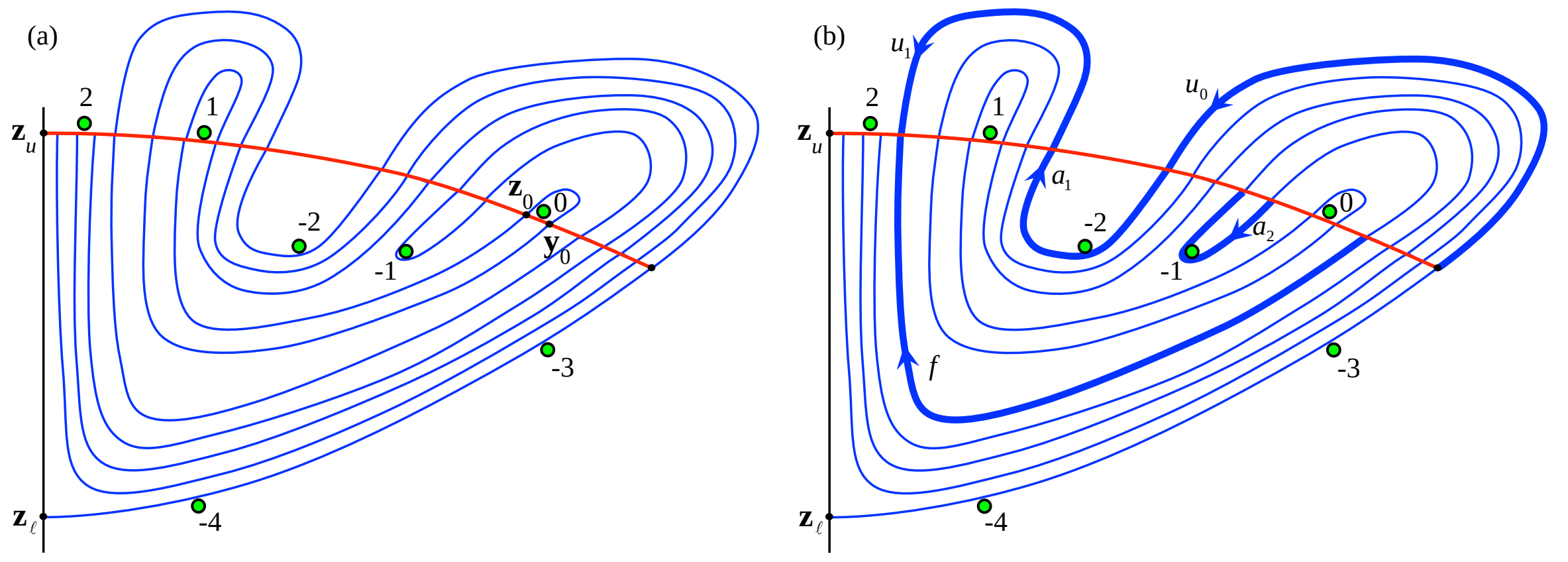

**Figure 5.** *Homotopy description of the Example 2 trellis. a) The pseudoneighbors $\mathbf{z}_0$ and $\mathbf{y}_0$ are shown. The holes associated with $\mathbf{z}_n$ and $\mathbf{y}_n$ are shown as the numbered green circles. b) The bridge classes $a_1$, $a_2$, $f$, $u_0$, and $u_1$ are shown in bold.*

We next introduce a geometric scheme to systematically identify and label bridge classes. The first step of this scheme is to divide phase space into regions based on:

1. the stable component of the trellis, i.e. the segment $W^S[\mathbf{z}_u, \mathbf{p}_0]$;
2. every bridge with an interior pseudoneighbor, i.e. a pseudoneighbor that is not an endpoint; (Note that pseudoneighbors $\mathbf{y}_{-1}$, $\mathbf{z}_{-1}$ and $\mathbf{y}_{-2}$, $\mathbf{z}_{-2}$ in Fig. 3b are interior to their bridges.)
3. every primary inert bridge whose endpoints form a pseudoneighbor pair;
4. the unstable segment $W^U[\mathbf{z}_\ell, \mathbf{p}_0]$.

See Fig. 6a. We call the partition of phase space formed by cutting along these curves the *primary division* of phase space. A bridge that lies on the boundary of a division region is assigned to a region based on a small distortion of the bridge. If a small distortion of the bridge can move it into the interior of region $D$ without passing through a hole, then that bridge is part of region $D$. For example, the bridge $W^U[\mathbf{p}_1, \mathbf{q}_0]$ in Fig. 6a lies in region I.

The construction of the primary division guarantees that every bridge in $W^U$ resides in one region, since bridges cannot intersect. Furthermore, no holes lie in the interior of a region, since all holes are attached to either an unstable segment of type (ii) or (iv) in the construction or to the stable segment (i). Thus, the intervals of $W^S[\mathbf{z}_u, \mathbf{p}_0]$ (shown in Fig. 6a) in which the endpoints of a bridge lie fully determine which holes the bridge wraps around, and hence to which bridge class the bridge belongs. We thus next examine how the stable segment $W^S[\mathbf{z}_u, \mathbf{p}_0]$ is divided up within the primary division.

The stable segment $W^S[\mathbf{z}_u, \mathbf{p}_0]$ can be divided according to either the regions interior to the resonance zone (the *inner stable* division) or those exterior to the resonance zone (the *outer stable* division). Figures 6b and 6c show these two divisions. The inner stable division consists of five stable intervals and the outer stable division three. The intervals are separated by half-filled circles, where the filled half of a circle indicates to which interval a point belongs. For example, the point $\mathbf{p}_1$ in Fig. 6b belongs to the interval from $\mathbf{p}_1$ to $\mathbf{z}_u$. This is because the bridge $W^U[\mathbf{p}_1, \mathbf{q}_0]$ lies in region I, as noted above. Similarly, the interval from $\mathbf{q}_0$ to $\mathbf{p}_1$ in

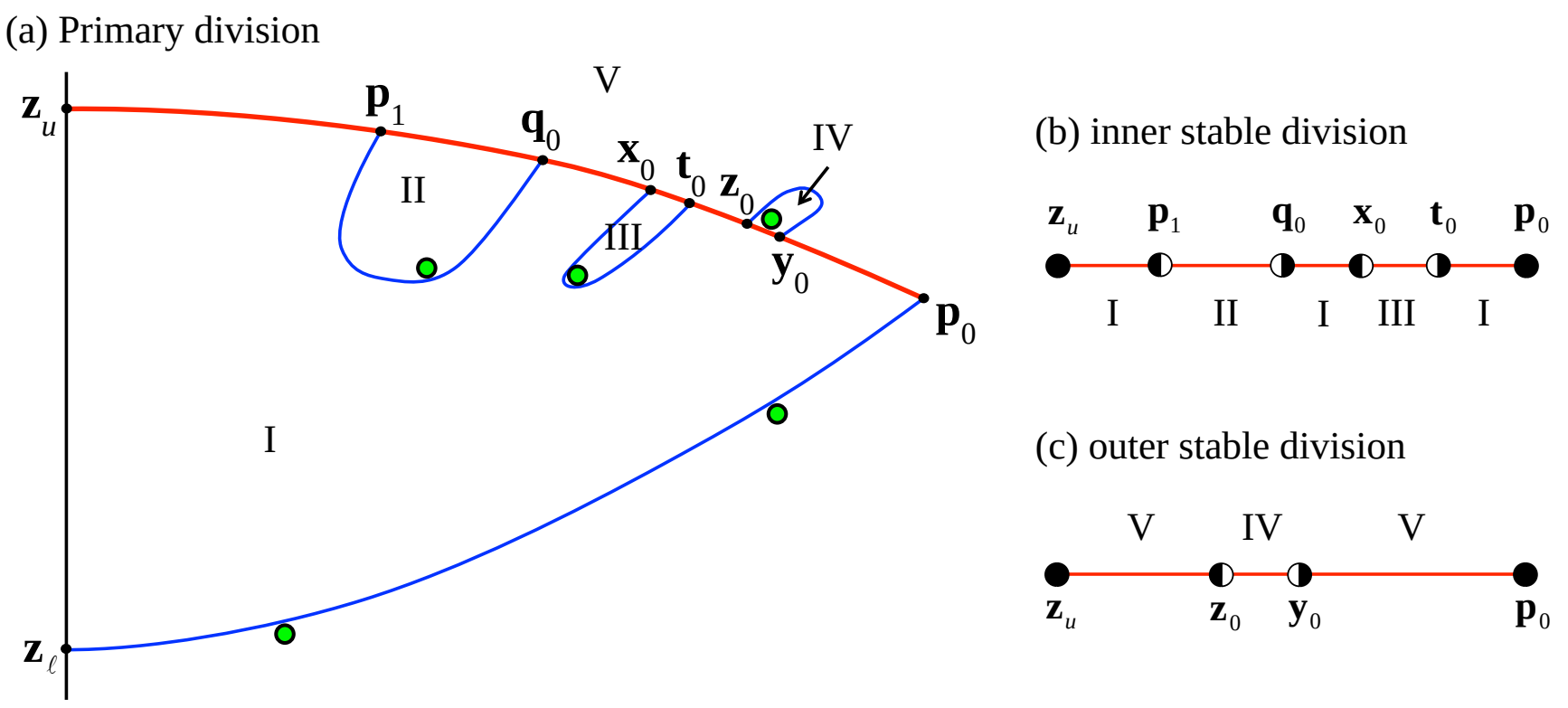

**Figure 6.** *Primary division for Example 2. a) The division of the 2D phase space. b) The 1D inner stable division of* $W^S[\mathbf{z}_u, \mathbf{p}_0]$*. c) The 1D outer stable division of* $W^S[\mathbf{z}_u, \mathbf{p}_0]$*.*

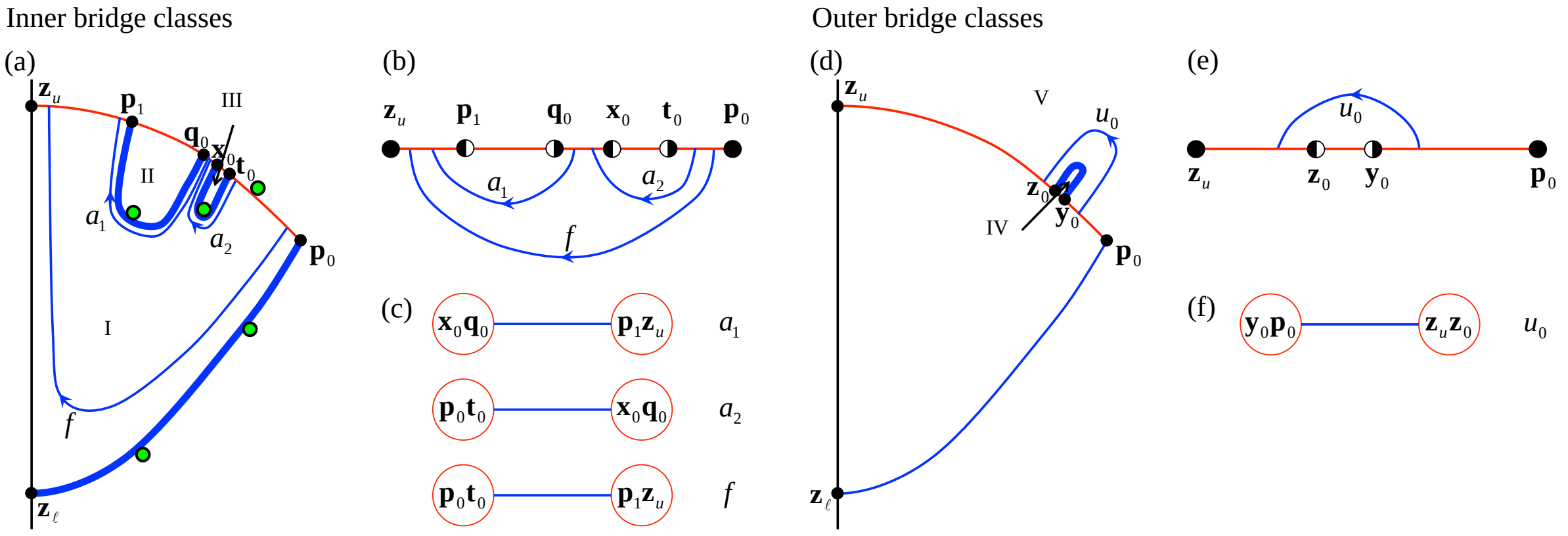

**Figure 7.** *Three representations of bridge classes for Example 2. a) Inner classes shown as directed curves within the primary division. Figure is squeezed in width relative to Fig. 6a. b) The curves from panel (a) are extracted to focus on the locations of their endpoints. c) Bridge-class "dumbbell" notation based on endpoint locations. Panels (d)-(f) repeat (a)-(c) for the outer classes.*

Fig. 6b is open on both ends.

Restricting attention to the active classes and the primary inert classes, a bridge class is now uniquely specified by the intervals of the stable division within which its endpoints lie, as shown in Fig. 7. Figure 7a shows the division of the inner zone with bold curves. The thin curves are the three bridge classes. The stable intervals connected by the inner classes are shown schematically in Fig. 7b. This suggests the new notation for bridge classes illustrated in Fig. 7c. Each bridge class is represented by a (blue) line connecting two (red) circles. The red circles indicate the stable intervals in which the endpoints of the bridge class lie. The points within the red circles are the endpoints of the stable intervals. For example, bridge class $a_1$ connects intervals $W^S[\mathbf{x}_0, \mathbf{q}_0]$ and $W^S[\mathbf{z}_u, \mathbf{p}_1]$. The other two inner bridge classes are also shown in Fig. 7c. The outer zone and its single bridge class are similarly analyzed in Figs. 7d-7f.

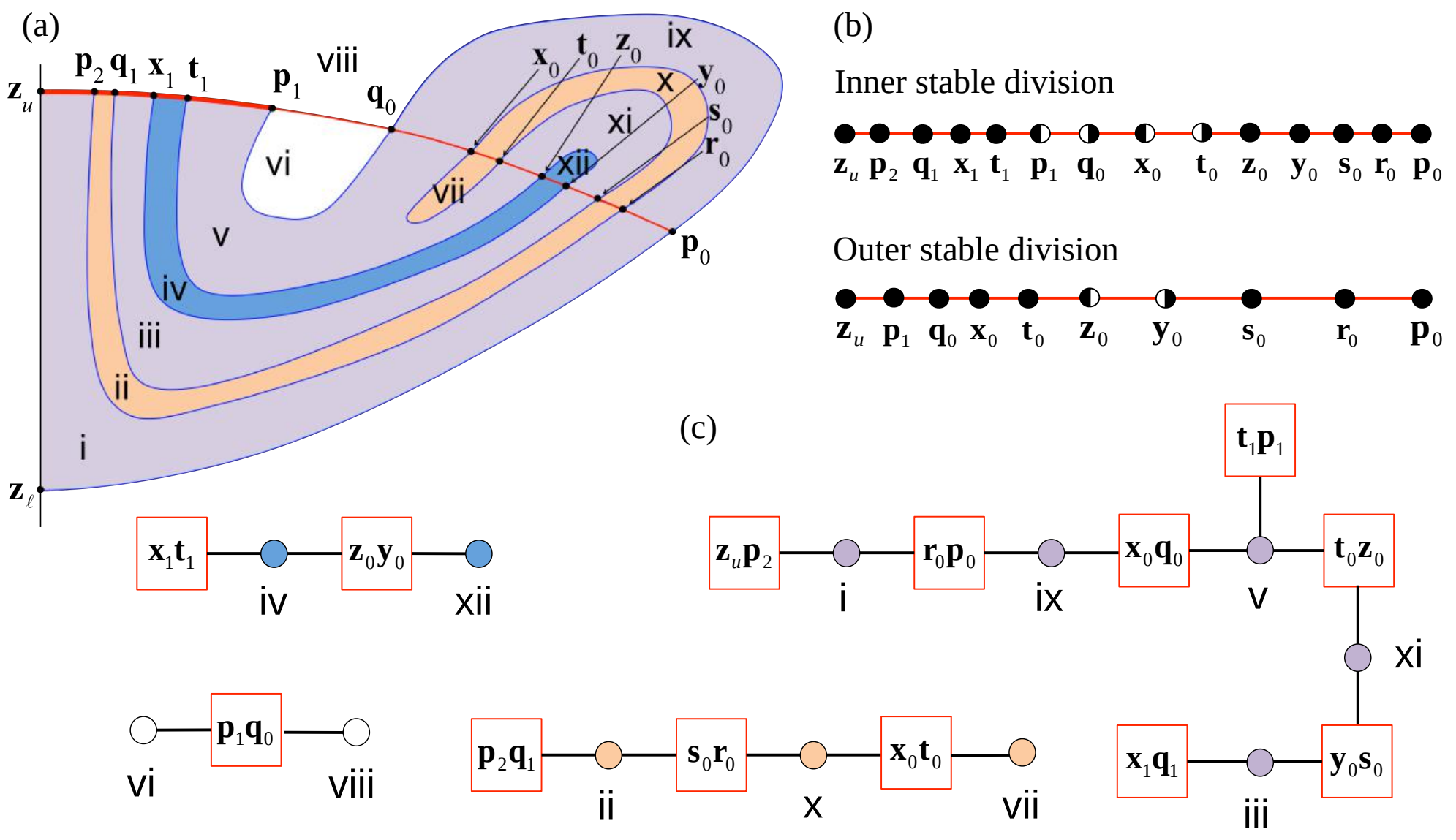

**Figure 8.** *Secondary division for Example 2. a) Division of the phase plane. b) The inner and outer divisions of the stable component. c) The connection graph for the regions of the secondary division.*

**2.2.5. Secondary division and bridge class iterates.** To determine the forward iterate of each bridge class, we construct a *secondary* division by cutting along the curves:

1. the stable component of the trellis, i.e. the segment $W^S[\mathbf{z}_u, \mathbf{p}_0]$;
2. the forward iterate of every bridge with a pseudoneighbor in its interior;
3. the forward iterate of the unstable segment $W^U[\mathbf{z}_\ell, \mathbf{p}_0]$.

The secondary division is thus obtained by iterating forward all the unstable segments from the primary division, except for the inert bridges. Note that all segments from the primary division are included in the secondary division. The regions of the secondary division are denoted by lower-case roman numerals in Fig. 8a. As with the primary division, the secondary division induces two separate divisions of the stable interval $W^S[\mathbf{z}_u, \mathbf{p}_0]$, shown in Fig. 8b.

We shall need to know how the different regions of the secondary division are connected to one another. Specifically, two regions will be said to be connected if they share a common boundary subinterval of the stable fundamental segment $W^S[\mathbf{p}_0, \mathbf{p}_1]$. For example, region v is connected to region ix, but not to region viii. The graph in Fig. 8c records how the secondary regions are connected to one another. Each region is denoted by a dot. Each boundary interval between connected regions is denoted by a red square with the boundary endpoints listed inside.

Recall that an (active) bridge class is specified by the intervals of the stable primary division in which its endpoints lie. Then the forward iterate of an (active) bridge class is specified by the intervals of the stable secondary division in which its endpoints lie. Fig. 9a shows how this applies to class $a_1$; the endpoints of $a_1$ lie in the intervals $W^S[\mathbf{x}_0, \mathbf{q}_0]$ and $W^S[\mathbf{p}_1, \mathbf{z}_u]$ of the stable primary division, so that the endpoints of $M(a_1)$ lie in the intervals $W^S[\mathbf{x}_1, \mathbf{q}_1]$ and $W^S[\mathbf{p}_2, \mathbf{z}_u]$ of the stable secondary division. Fig. 9b shows these intervals as bold red segments within the secondary division. $M(a_1)$ must connect these intervals by a curve in the secondary division that (i) does not cross an unstable (blue) segment and (ii) may only cross the stable (red) segment along the fundamental interval $W^S[\mathbf{p}_0, \mathbf{p}_1)$. The unique way to do this is illustrated by the green curve in Fig. 9b. A more systematic approach to constructing this path is shown in Fig. 9c using the connection graph. The squares representing the beginning and ending intervals $W^S[\mathbf{x}_1, \mathbf{q}_1]$ and $W^S[\mathbf{p}_2, \mathbf{z}_u]$ are shaded green. Because the connection graph is a tree, there is a unique shortest path (in green) connecting these green squares.

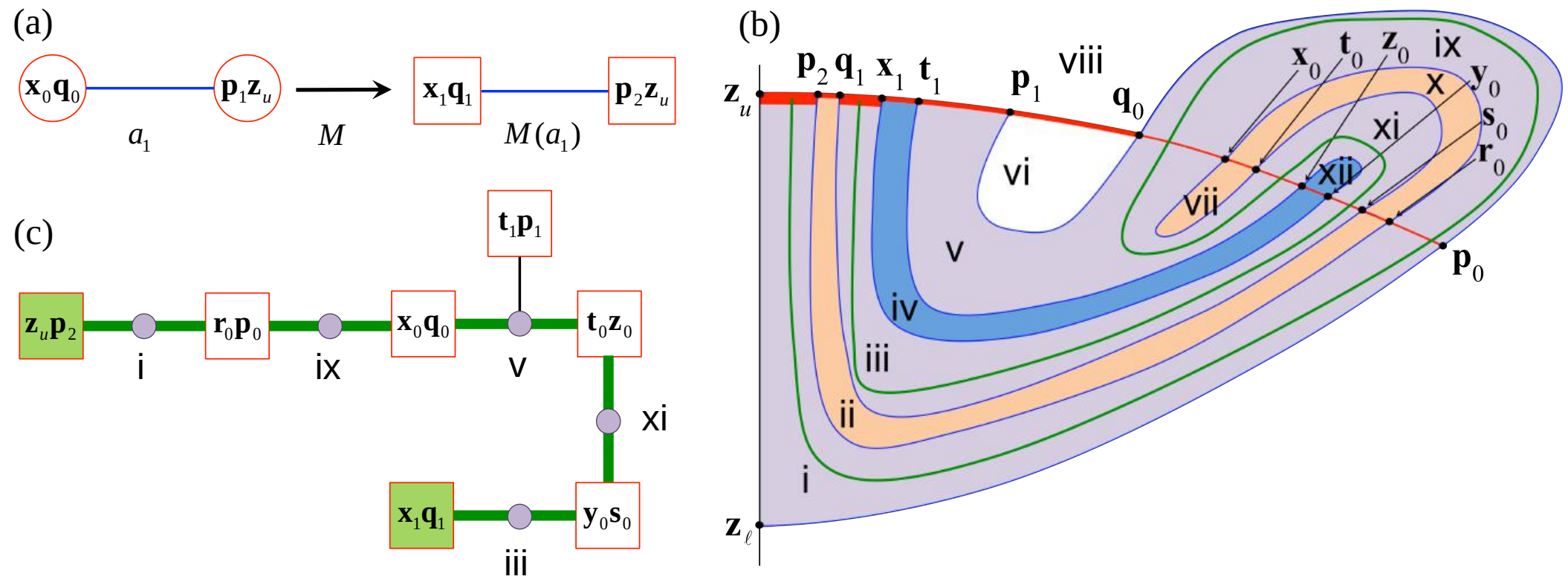

**Figure 9.** *Constructing the forward iterate of $a_1$. a) The terminal stable segments of $a_1$ (red circles) are mapped forward to the stable segments in the red boxes. b) The green curve connects the bold red segments (corresponding to the boxes in part a) without intersecting the blue unstable curves. c) The green line connects the two shaded green boxes (the same boxes as in part a) within the connection graph.*

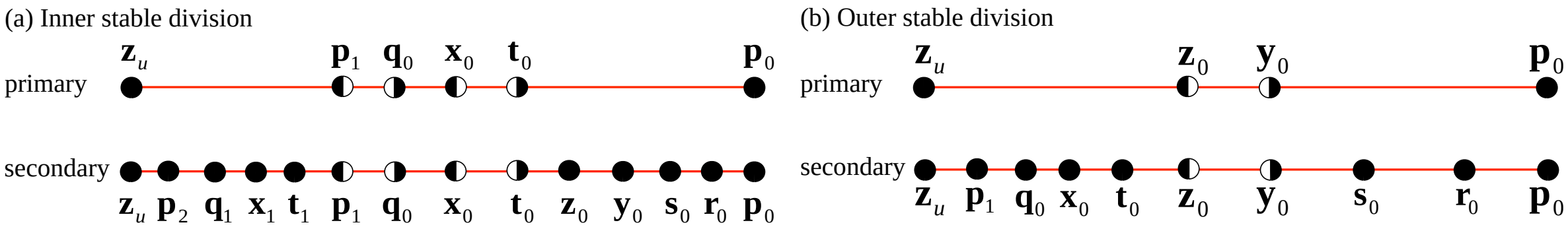

**Figure 10.** *All intervals of the secondary stable division are subintervals of the primary stable division. The inclusion map is different for the inner stable divisions (panel a) and the outer stable divisions (panel b).*

We next convert the green path in Fig. 9c into a sequence of bridge classes. In moving from one region of the secondary division to the next along the green path, we intersect an interval of the stable secondary division. For example, going from region i to region ix, we cross $W^S[\mathbf{r}_0, \mathbf{p}_0]$. Note, each interval in the secondary stable division is a subinterval of an interval in the primary stable division, as shown in Fig. 10. Of course, the inclusion map taking an interval of the secondary stable division into the primary stable division depends on which side of the stable manifold, inner versus outer, one is working with. For example $W^S[\mathbf{r}_0, \mathbf{p}_0]$ embeds in the interval $W^S[\mathbf{t}_0, \mathbf{p}_0]$ of the inner stable division but in the interval $W^S[\mathbf{y}_0, \mathbf{p}_0]$ of the outer stable division. We can now apply this embedding to the green path as shown in Fig. 11. The green path is reproduced at the bottom of Fig. 11a, with solid segments in the inner zone and dashed segments in the outer zone. Each stable interval from the inner secondary division is embedded into a stable interval from the *inner* primary division, according to Fig. 10a. Likewise, Fig. 11b repeats this process for the segments of the green curve in the outer zone (now shown as solid) using the embedding into stable intervals from the *outer* primary division, according to Fig. 10b. The final result is given in Fig. 11c, showing that the original class $a_1$ maps to a sequence of five bridge-classes that alternate between inner and outer classes. Below each "dumbbell" representation of a bridge class, we have written the corresponding symbol for the class, given in Figs. 7c and 7f. $M(a_1)$ can thus be expressed as the product $f^{-1}u_0{a_2}^{-1}{u_0}^{-1}f$. $M(a_2)$ and $M(f)$ are obtained using the same approach as $M(a_1)$, yielding the three symbolic equations

$$M(a_1) = f^{-1}u_0{a_2}^{-1}{u_0}^{-1}f, \tag{5}$$

$$M(a_2) = f^{-1}{u_0}^{-1}f, \tag{6}$$

$$M(f) = f^{-1}{u_0}^{-1}a_1. \tag{7}$$

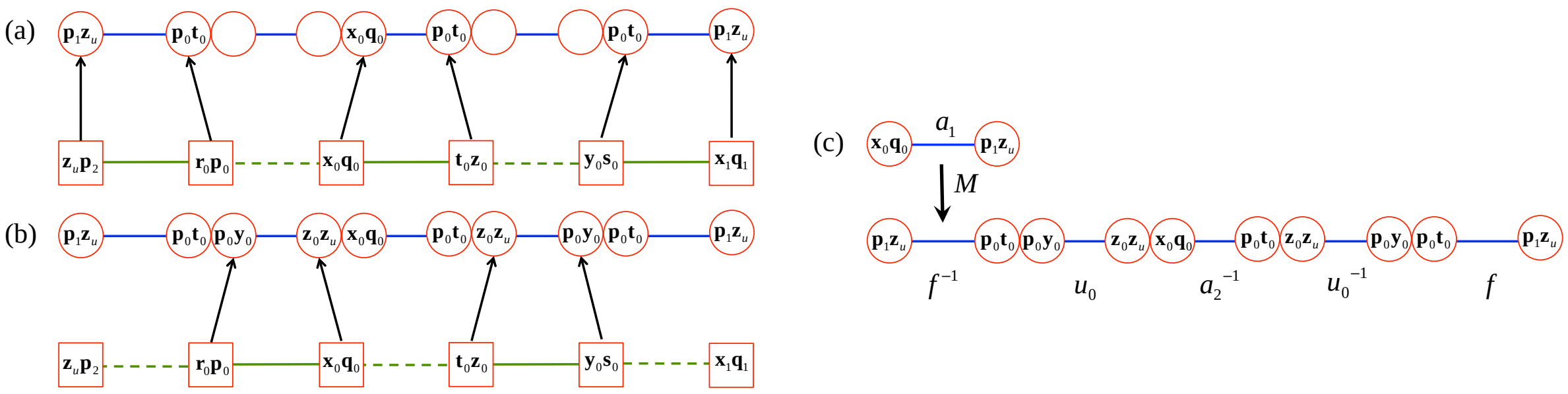

**Figure 11.** *a) The inclusion of segments from the inner secondary division (squares) into the inner primary division (circles) is shown by the arrows. b) The inclusion of segments from the* outer *secondary division into the* outer *primary division. c) The iterate of* $a_1$ *is summarized.*

**2.2.6. Transition matrix and topological entropy.** The *transition matrix* $\mathsf{T}$ for the above symbolic dynamics has components $T_{ij}$ that record the number of times the bridge class of row $i$ appears in the forward iterate of the bridge class of column $j$.

$$\mathsf{T} = \begin{array}{c} \\ a_1 \\ a_2 \\ f \end{array} \begin{array}{c} \begin{array}{ccc} a_1 & a_2 & f \end{array} \\ \begin{pmatrix} 0 & 0 & 1 \\ 1 & 0 & 0 \\ 2 & 2 & 1 \end{pmatrix} \end{array} \tag{8}$$

Note that we need not include the inert classes, having the trivial dynamics Eq. (4). The transition matrix certainly does not include the full information of Eqs. (5)–(7). In particular the ordering of symbols and the inverses on symbols are lost. Nevertheless, the transition matrix is a useful distillation of the full equations. For example, the topological entropy of Eqs. (5)–(7) is the natural logarithm of the largest eigenvalue of $\mathsf{T}$,

$$h_{\text{top}} = \ln 2.2695. \tag{9}$$

The homotopic lobe dynamics algorithm followed here can be summarized by the outline below.

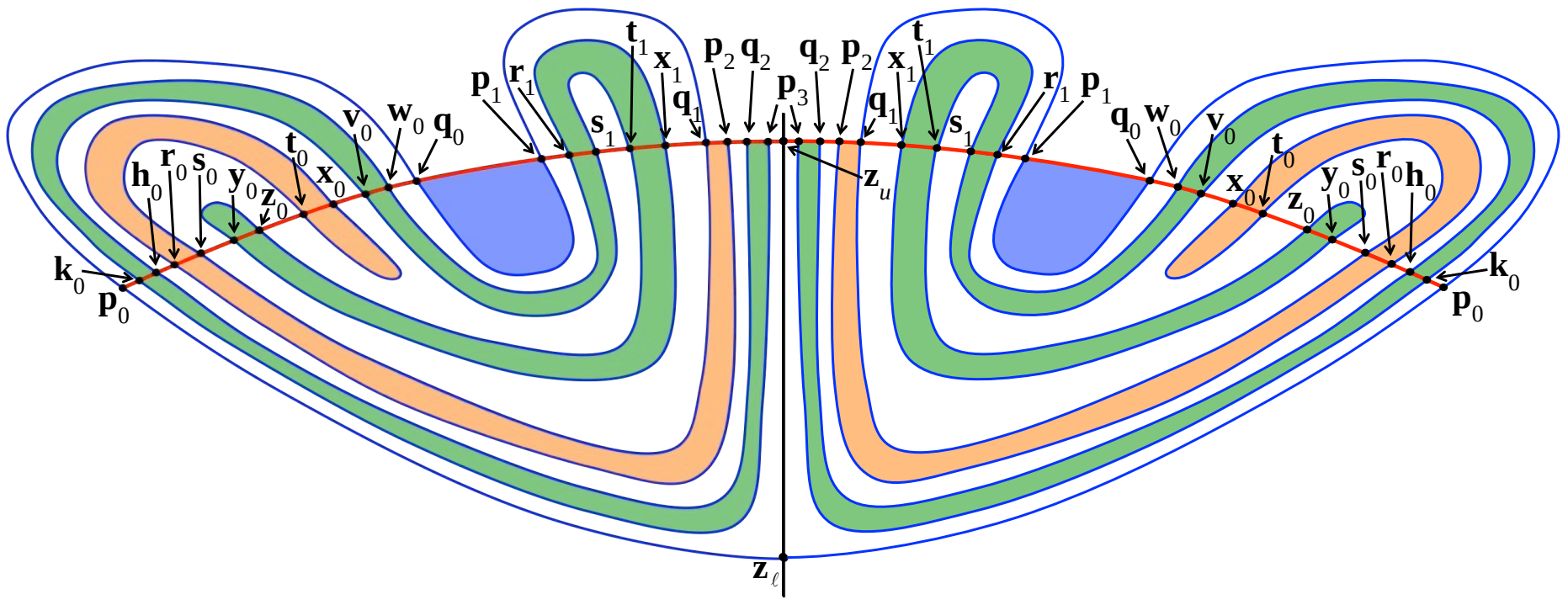

**Figure 12.** *Cross section of the rotationally symmetric 3D trellis for Example 2.*

1. Begin with a trellis—finite pieces of stable and unstable manifolds.
2. Examine forward/backward escape-time plots on fundamental segments and identify pseudoneighbors.
3. Punch holes adjacent to pseudoneighbors.
4. Construct primary division and use it to specify the bridge classes.
5. Construct secondary division and use its connection graph to construct iterates of bridge classes.

Once the iterates of bridge classes are known, then a transition matrix can be defined, and the topological entropy, or other measures of complexity, can be computed. As we shall see, this outline applies to the 3D HLD procedure as well.

**2.3. 3D geometry of tangles with an axis of symmetry.** We obtain a trellis in 3D by rotating the planar trellis in Fig. 3a about the axis joining $\mathbf{z}_\ell$ to $\mathbf{z}_u$. A cross-section of this trellis is shown in Fig. 12. The stable and unstable manifolds $W^S$ and $W^U$ are now 2D surfaces, and the heteroclinic intersections between these surfaces are 1D circles. We denote the stable component of the trellis, i.e. the finite piece of the unstable manifold, by $T^S$. It is the cap whose boundary is the single circle $\mathbf{p}_0$. (For continuity of exposition, we use the same bold notation $\mathbf{p}_0$ for the entire intersection curve as we had used for a single intersection point.) There is a corresponding unstable cap containing $\mathbf{z}_\ell$ and also bounded by $\mathbf{p}_0$. Together these two caps enclose the resonance zone, which is now a 3D volume.

The stable cap $T^S$ cuts the unstable manifold $W^U$ into domains bounded by intersection circles. We denote the unstable domain bounded by circles $\mathbf{a}$, $\mathbf{b}$, $\mathbf{c}$, ..., by $W^U[\mathbf{a}, \mathbf{b}, \mathbf{c}, ...]$. The bottom cap containing $\mathbf{z}_\ell$ is thus denoted $W^U[\mathbf{p}_0]$. The top domain $W^U[\mathbf{p}_0, \mathbf{q}_0]$ takes the form of a "bundt-cake". (See Fig. 13b.) As for 1D manifolds, we define a 2D bridge as a domain of $W^U$ whose boundary circles lie on $T^S$ and that does not otherwise intersect $T^S$. In Fig. 12, one bridge forms a cap (Fig. 13a), but all other bridges are bundt-cakes (Fig. 13b). However, for general 2D tangles with broken rotational symmetry, bridges can have any number of boundary circles forming different geometries, such as a "macaroni" shape (Fig. 13c, seen in Example 5) or a "tridge" (Figs. 13d-13g), so-called because of its three boundary circles [1].

[1]The term "tridge" is taken from the eponymous structure in Midland, MI.

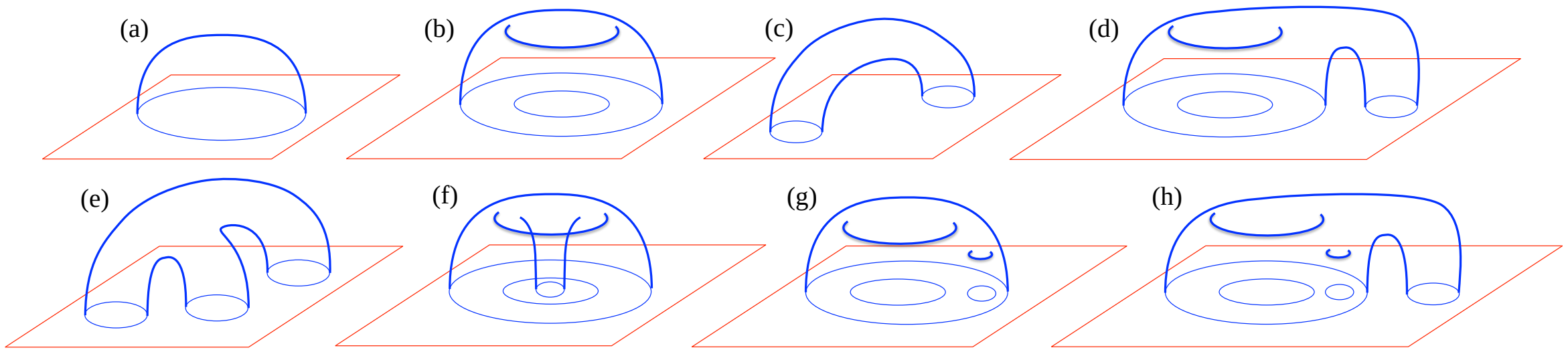

**Figure 13.** *An assortment of different bridge shapes. (a) cap; (b) bundt-cake; (c) macaroni; (d)-(g) tridges: each tridge differs by the nesting of its boundary circles; (h) a bridge with four boundary circles.*

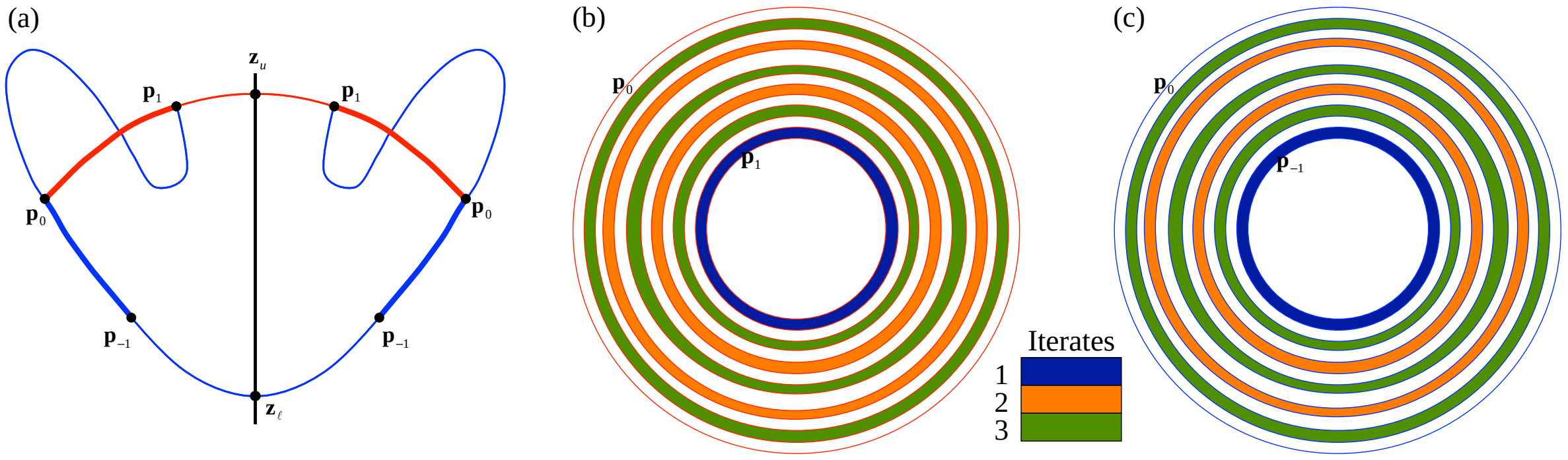

**Figure 14.** *a) Cross-sections of the 2D stable and unstable fundamental annuli are shown as bold red and blue segments, respectively, for the rotationally invariant 3D version of Example 2. Panels b) and c) show the backward and forward escape-time plots.*

The four tridges in Fig. 13 differ by the manner in which their boundary circles are nested. A tridge is first seen in Example 3. Figure 13h shows one of many bridges with four boundary circles.

Because the manifolds in Fig. 4a are rotated about the vertical axis, the 1D fundamental segments $W^S[\mathbf{p}_0, \mathbf{p}_1]$ and $W^U[\mathbf{p}_0, \mathbf{p}_{-1}]$ sweep out fundamental *annuli*. The intersections of the annuli with a 2D plane are shown as the bold segments in Fig. 14a. We use the same notation $W^S[\mathbf{p}_0, \mathbf{p}_1]$ and $W^U[\mathbf{p}_0, \mathbf{p}_{-1}]$ for the stable and unstable 2D fundamental annuli as we did for the fundamental segments, though now $\mathbf{p}_0$ and $\mathbf{p}_1$ denote curves rather than points. For the 3D map, an escape-time plot is generated by counting the number of forward (backward) iterates it takes an arbitrary point on the unstable (stable) fundamental annulus to escape the resonance zone. Fig. 14b illustrates the forward escape-time plot, where the color denotes the number of iterates to escape. Similarly, Fig. 14c illustrates the backward escape-time plot. Compare these plots to Fig. 4b and Fig. 4c.

As in 2D, the escape-time plots can be used to locate pairs of pseudoneighbors. (We show explicitly how to do this in the next section.) Now, however, the pseudoneighbors are 1D objects, and 1D holes are then punched next to them. We can imagine these holes physically as *rings* embedded in the 3D phase space, as shown in Fig. 15.

Just as we imagined rods stirring the 2D phase space (Fig. 2d), we can imagine the rings animated in time stirring the 3D phase space. Under this animation the rings can shrink or grow in time and one ring can move through the center of another ring. The 2D stirring protocol in Fig. 2d becomes the 3D stirring protocol illustrated in Fig. 15b. Though the analysis technique in this paper does not make explicit use of the animation of rings, it is a convenient physical picture to have in mind as one seeks to understand how 2D surfaces are stretched and folded under the dynamics.

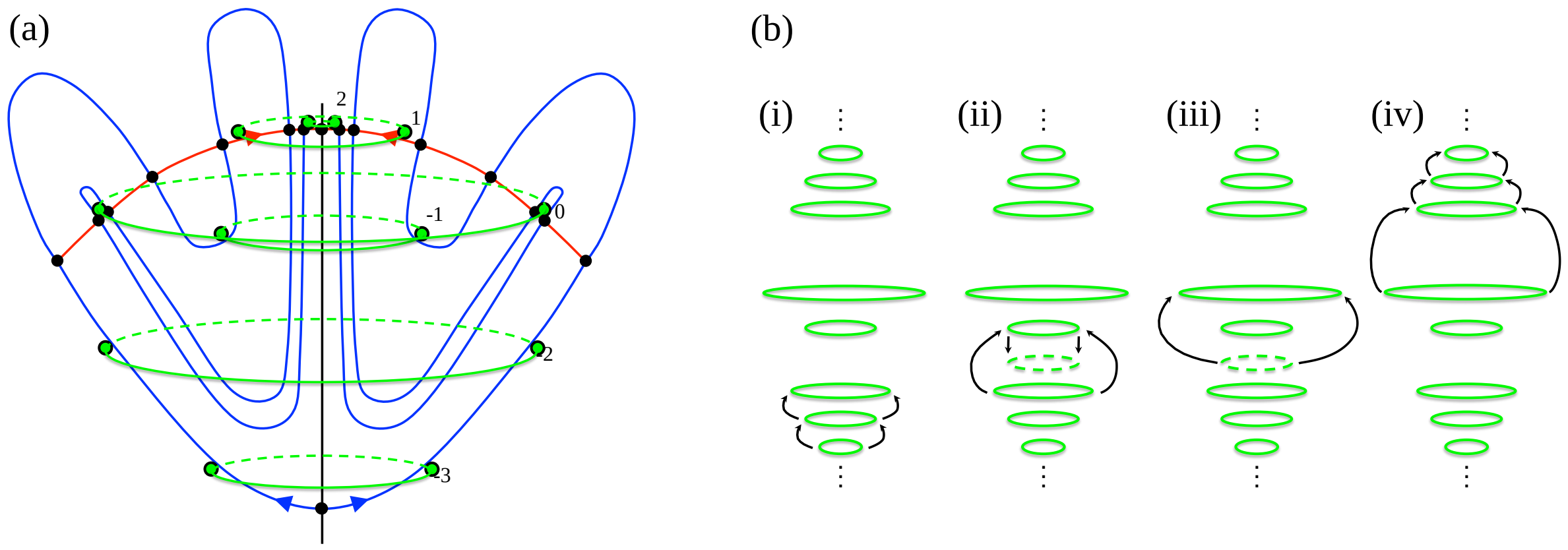

**Figure 15.** *a) Cross section of the rotationally symmetric trellis in 3D modelled on the 2D trellis in Fig. 2. The holes in Fig. 2b become rings (green) in 3D. b) The 2D stirring protocol in Fig. 2b becomes the 3D stirring protocol described by the four steps: (i) An infinite number of rings moves up from the bottom, with each ring advancing to the position of the next ring. (ii) One ring expands and moves up to the position of the next ring. The next ring simultaneously move down to a resting position. (iii) The ring in the resting position expands and leapfrogs over the ring above it to occupy the middle position. (iv) The middle ring and all rings above advance forward.*

## 3. Maps without an axis of symmetry.

**3.1. Example 3: A 3D horseshoe with variable overshoot.** As we further extend our method to 3D, we turn our attention to examples without an axis of symmetry. We follow the outline at the end of Sect. 2.2

**3.1.1. Trellis.** In the cross-section view of the trellis in Fig. 16a, we see the geometry of a complete horseshoe on the right and a horseshoe with overshoot on the left. (Compare to Figs. 2a and 3a.) To form a complete picture of the topology of the trellis, Fig. 16b shows the intersection curves between $T^U$ and $T^S$ on the stable cap. Notice that the intersection curves $\mathbf{t}_0$, $\mathbf{t}_1$, and $\mathbf{x}_0$ do not enclose the fixed point $\mathbf{z}_u$ within $T^S$. This clearly illustrates the broken symmetry. Note, some bridges now have *more* than two boundary circles, for example the tridge $W^U[\mathbf{s}_0, \mathbf{t}_0, \mathbf{q}_1]$.

**3.1.2. Escape-time plots and pseudoneighbors.** We again begin our analysis by locating pseudoneighbors. Recall that in 3D the fundamental segments are replaced by fundamental *annuli*, as shown in Fig. 17. In the forward escape-time plot (Fig. 17b), the blue domain escapes the resonance zone in one iterate, the orange domains escape in two iterates, and the green domains escape in three iterates. The shaded domains in the backward escape-time plot (Fig. 17a) are similarly color-coded. These escape-time plots can be determined entirely from the data in Fig. 16, similar to how Figs. 4b and 4c were obtained from Figs. 3a and 3b. In particular, Fig. 17a is essentially just that part of Fig. 16b between $\mathbf{p}_0$ and $\mathbf{p}_1$, shaded according to the lobes in Fig. 16a.

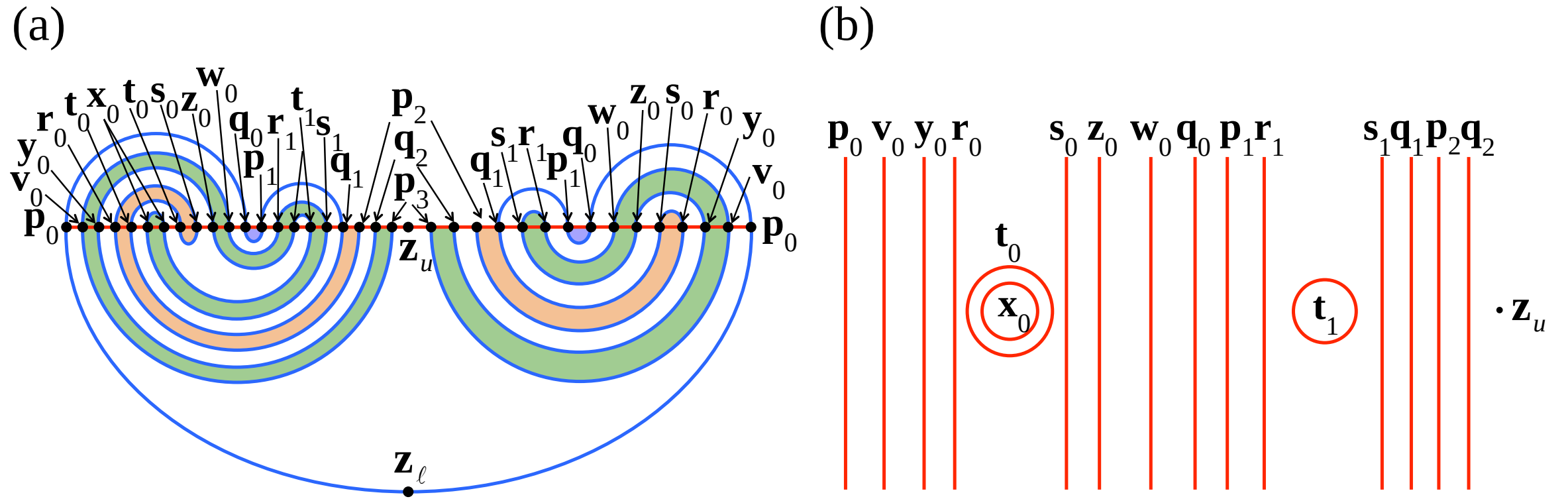

**Figure 16.** *The trellis for Example 3. (a) A vertical cross section. A more schematic view of the trellis is given here, compared to Fig. 12. (b) The top-down view of the stable cap. For visualization purposes, the stable cap is not represented as a circle, as in Figs. 14a and 14b. To reconstruct circles, the top and bottom points of each vertical segment in (b) should be identified.*

Figure 17b requires more mental gymnastics to construct. One approach would be to construct the backward iterates of the stable manifold, as in Fig. 3b, and then read off the forward escape-time plot from the intersections of the stable manifold with the unstable cap $W^U[\mathbf{p}_0]$. A more direct approach, however, is to simply consider how $W^U[\mathbf{p}_{-1}, \mathbf{p}_0]$ maps forward. The first iterate is $W^U[\mathbf{p}_0, \mathbf{p}_1]$, which from Fig. 16 is seen to intersect $W^S[\mathbf{p}_0]$ at $\mathbf{q}_0$, forming the bridge $W^U[\mathbf{p}_0, \mathbf{q}_0]$ that has escaped on the first iterate. This bridge produces the blue escape domain on the far right of Fig. 17b. Iterating the inner bridge $W^U[\mathbf{p}_1, \mathbf{q}_0]$ forward to $W^U[\mathbf{p}_2, \mathbf{q}_1]$, we obtain the new intersections $\mathbf{s}_0$, $\mathbf{r}_0$, and $\mathbf{t}_0$. These define the escaped bridges $W^U[\mathbf{s}_0, \mathbf{r}_0]$ and $W^U[\mathbf{t}_0]$, which produce the two orange escape domains in Fig. 17b. Continuing in this manner, mapping the inner bridges $W^U[\mathbf{s}_0, \mathbf{t}_0, \mathbf{q}_1]$ and $W^U[\mathbf{r}_0, \mathbf{p}_2]$ forward we obtain the green escape domains in Fig. 17b. Comparing Figs. 17a and 17b to Figs. 14b and 14c, not only has the rotational symmetry been broken, but some escape domains, or gaps between escape domains (the white zones), have more than two boundary components. For example, the orange domain in Fig. 17b has three boundary components, $\mathbf{r}_0$, $\mathbf{s}_0$, and $\mathbf{t}_0$.

We now generalize the definition of pseudoneighbors to 3D maps. Two heteroclinic intersection curves that have iterates adjacent to each other in both the forward and backward escape-time plots are said to form a pseudoneighbor pair. This is demonstrated in Figs. 18a and 18b. For example, a line can be drawn between $\mathbf{s}_n$ and $\mathbf{r}_n$ without crossing any other intersection curve; this is true within both the stable and unstable fundamental annuli. Thus, $\mathbf{s}_n$ and $\mathbf{r}_n$ form a pseudoneighbor pair. Similarly, $\mathbf{s}_n$ and $\mathbf{t}_n$ form a pseudoneighbor pair. Furthermore, notice that the intersection curve $\mathbf{x}_n$ by itself encloses a single domain of $W^S$ and $W^U$ in Figs. 18a and 18b. For this reason, $\mathbf{x}_n$ is said to be a pseudoneighbor with itself, demonstrating a new possibility in 3D.

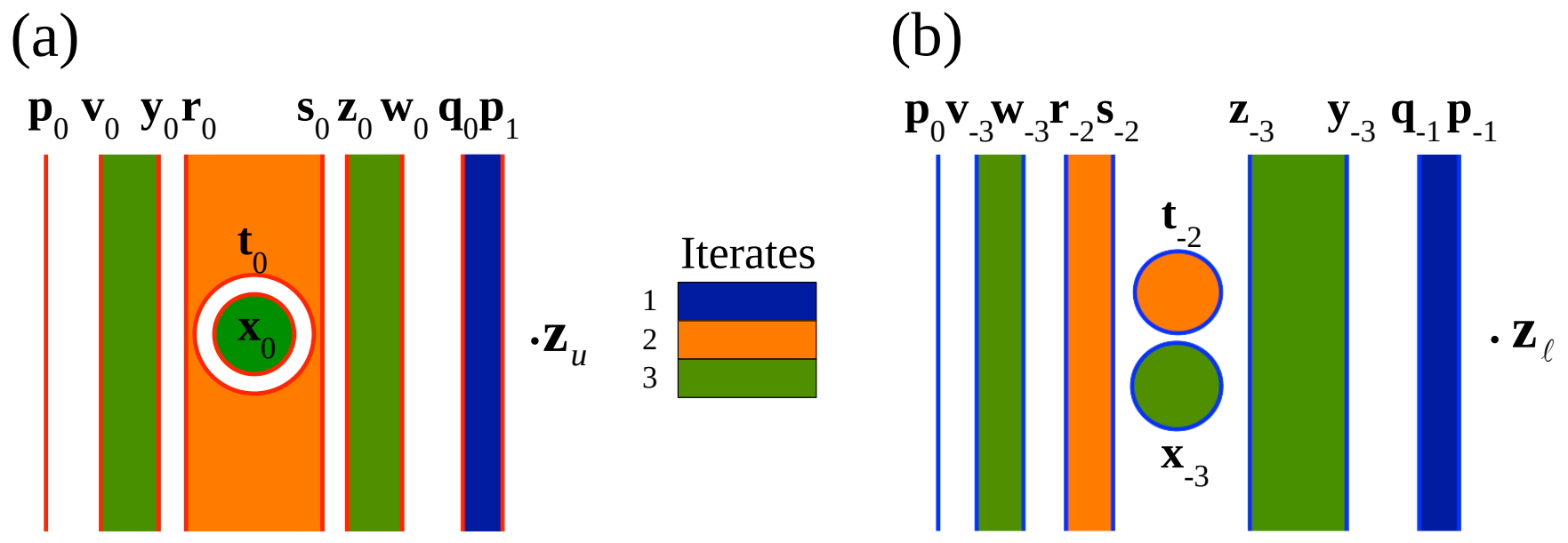

**Figure 17.** *(a) Backward and (b) forward escape-time plots for Example 3.*

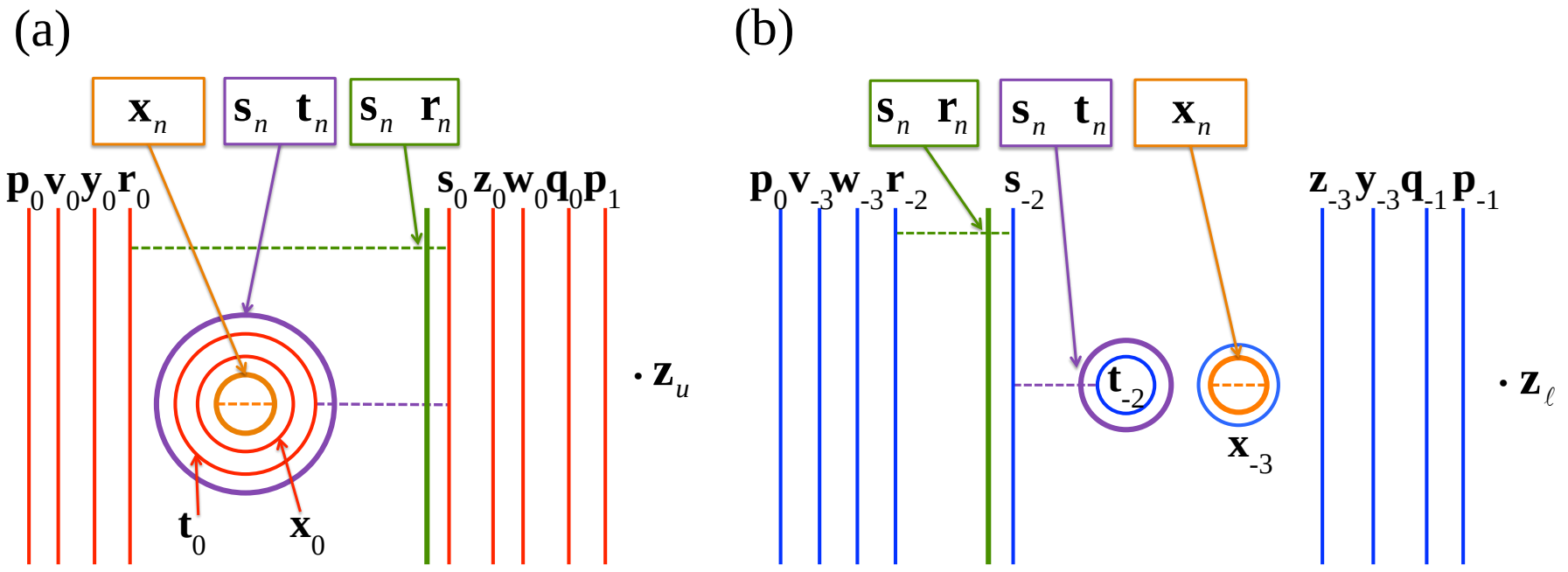

**Figure 18.** *Determination of the pseudoneighbors. Panels (a) and (b) show heteroclinic intersection curves on the stable and unstable fundamental annuli, respectively. The dashed lines connect pseudoneighbor pairs or, in the case of $\mathbf{x}_n$, connect a self-pseudoneighbor to itself. The green, purple, and orange curves represent the placement of holes.*

**3.1.3. Holes.** We next discuss the placement of holes adjacent to pseudoneighbors in the 3D phase space. We select one curve from each pseudoneighbor pair and remove a topological circle infinitesimally close to the curve. This circle is perturbed along the stable direction from the pseudoneighbor curve toward its partner. For example, the green hole in Fig. 18a is infinitesimally close to $\mathbf{s}_0$ but perturbed outward toward its partner $\mathbf{r}_0$. Thus, in this figure, the hole cuts across the line joining $\mathbf{s}_0$ to $\mathbf{r}_0$. Similarly, the hole is perturbed along the unstable direction from the pseudoneighbor curve toward its partner. Note the placement of the green hole in Fig. 18b between $\mathbf{s}_{-2}$ and $\mathbf{r}_{-2}$. For an intersection that forms a pseudoneighbor with itself, e.g. $\mathbf{x}_n$ in Fig. 18, the hole is perturbed inward, into the disc bounded by the pseudoneighbor curve.

**3.1.4. Primary division and bridge classes.** With pseudoneighbors located and holes placed, we construct the primary division of phase space, shown in Fig. 19a. Analogous to the 2D example, the primary division is obtained by cutting phase space using:

1. the stable component of the trellis, i.e. the cap $T^S = W^S[\mathbf{p}_0]$;
2. any bridge that includes a pseudoneighbor in its interior;
3. any bridge with a boundary circle that is a primary inert pseudoneighbor, i.e. the first iterate of a pseudoneighbor to land on $T^S$, and for which the corresponding hole is nudged toward the interior of the bridge. (This includes all primary inert bridges whose boundary circles form a pseudoneighbor pair.)

Note that unlike the 2D case, we do not need to treat the unstable domain that includes the lower fixed point $\mathbf{z}_\ell$ separately, since in 3D this domain is a bridge (the cap $W^U[\mathbf{p}_0]$) that includes pseudoneighbors in its interior (all pseudoneighbors map backward toward $\mathbf{z}_\ell$), and hence falls under condition 2 above. In Fig. 19a, there are only three regions of the interior division, despite the fact that there are six regions in the cross-section. This is because the two regions labelled I in the cross section are connected in the 3D phase space, and similarly for regions II and III. Connected regions are colored the same color for ease of visualization. To explain the structure of the regions better, Fig. 19b focuses on just the bridges used to divide the phase space. Each color of curve represents a single bridge, e.g. the three cyan curves are all part of the same 2D bridge. Each bridge is labelled by a letter that refers to the bridges in Fig. 13. Thus, the cyan bridge, labelled d, has the structure of Fig. 13d.

As in the 2D case, the stable trellis $T^S$ may be divided according to either the division of phase space inside the resonance zone or the division outside the resonance zone. These inner and outer stable divisions are shown in Figs. 19c and 19d, respectively. As in the 2D case, bridge classes are distinguished by how they wrap around the holes. This may be specified by which domains of the stable division their boundary circles lie within as well as the homotopy class of the boundary circle within that stable division domain. (We ignore the orientation of the homotopy class here.) We call these homotopy classes *boundary classes*. We illustrate boundary classes using the green curves in Fig. 19e and Fig. 19f. The boundary classes in Fig. 19e describe the boundary circles of inner bridge classes, whereas the boundary classes in Fig. 19f describe the boundary circles of outer bridge classes. The reader can verify that each of the intersection curves in Fig. 16b generates a green boundary class in both Fig. 19e and Fig. 19f. On the other hand, the boundary class $F$ in Fig. 19e is not represented by an intersection in the trellis, but rather by $\mathbf{t}_1$, the iterate of the trellis intersection $\mathbf{t}_0$. Be careful not to confuse the boundary classes here with the bridge classes for the 2D analysis of Sect. 2. The classes here describe curves within the (outer or inner) primary division of the stable cap. The classes in Sect. 2 describe 1D bridges in the punctured phase space.

Though not obvious, it turns out that bridge classes are uniquely determined by their boundary classes. We introduce the double-bracket notation $[\![X, Y, ...]\!]$ for the bridge class with boundary classes $X, Y, ...$. Note that the order of the boundary classes within the double bracket does not matter; that is, different orderings correspond to the same bridge class, i.e. $[\![X, Y, Z]\!] = [\![Z, Y, X]\!]$. Bridge classes for the trellis in Fig. 16 are

$$\text{Inner:} \quad [\![A]\!], \quad [\![C, D]\!], \quad [\![A, E]\!], \quad [\![B, C, D]\!], \quad [\![B, C, D, F]\!], \quad [\![A, E, E]\!], \tag{10}$$

$$\text{Outer:} \quad [\![H]\!], \quad [\![G, I]\!]. \tag{11}$$

All of these classes have a representative bridge in the trellis except for $[\![A, E, E]\!]$. This class is only "discovered" later by mapping class $[\![B, C, D, F]\!]$ forward. (See Fig. 26.) We also represent the bridge classes using a modification of the graphical notation used in Figs. 7c and 7f for the 2D case. Bridge classes for the trellis in Fig. 16 are shown in Fig. 20. As before, each bridge class is represented by blue lines connecting red circles to a central dot. The dot and attached lines are interpreted as the unstable piece of $T^U$ and the red circles are interpreted as the boundary circles, with their boundary class recorded inside. In the case of 2D maps, each central dot is connected to exactly two red circles, due to the fact that every bridge has exactly two endpoints in 2D. In the case of 3D maps, the central dot can be connected to any number of red circles, since we may have one or more boundary circles on a bridge in 3D. Indeed in Fig. 20, we see bridge classes with one, two, three, or four boundary classes. Note that the placement of the boundary classes around the central blue dot is arbitrary, i.e. two representations with the same boundary classes in different orders still represent the same bridge class.

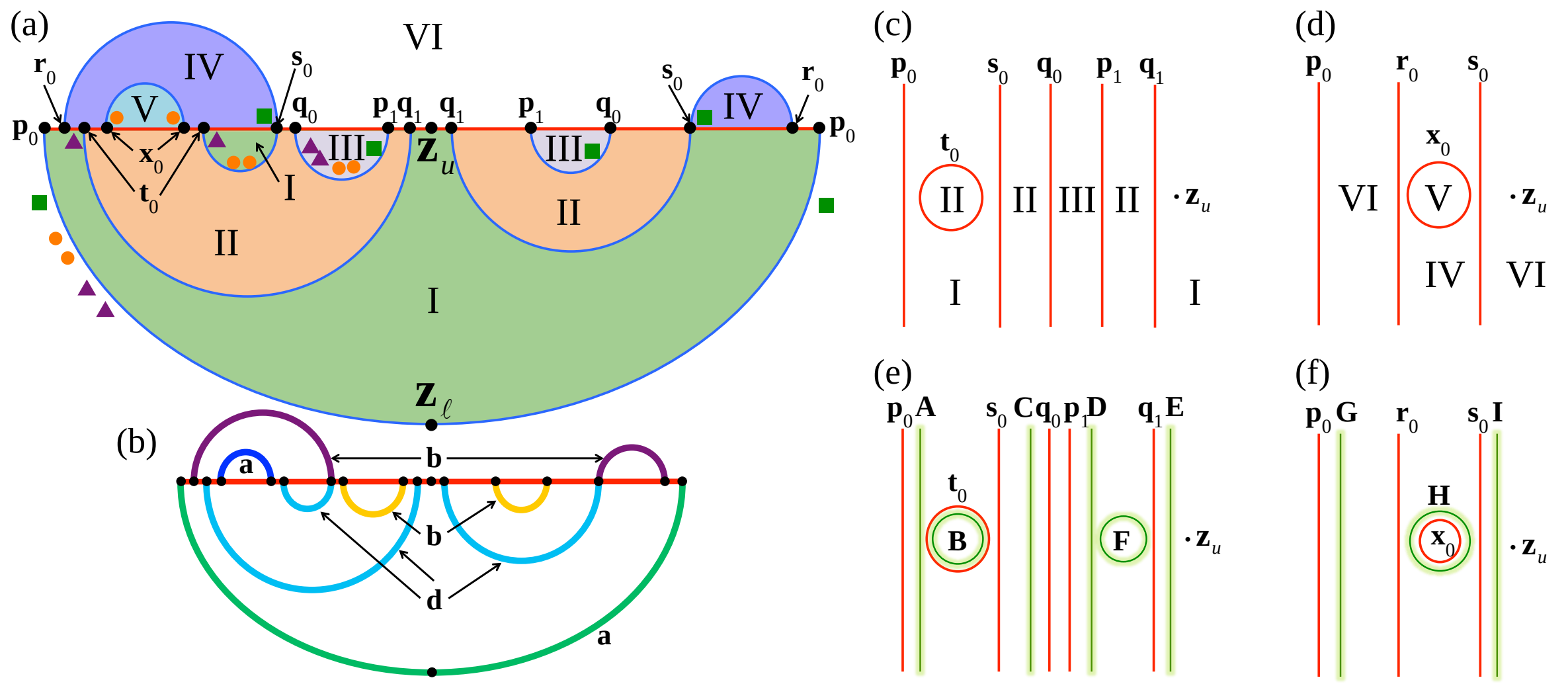

**Figure 19.** *Primary division for Example 3 (Fig. 16). a) The division of phase space. b) Each bridge from part (a) is color coded and labelled by the corresponding subfigure of Fig. 13. c) The inner stable division. d) The outer stable division. In (e) and (f) the boundary classes are shown in green.*

The topology of the classes in Fig. 20 can be connected to the images in Fig. 13 based on how the boundary circles are nested within one another. Classes $[\![A]\!]$ and $[\![H]\!]$ are obviously just caps (Fig. 13a). Class $[\![C, D]\!]$ is a bundt cake (Fig. 13b), since any curve of class $D$ is nested inside a curve of class $C$, as seen from Fig. 19e. By the same logic, $[\![A, E]\!]$ and $[\![G, I]\!]$ are also both bundt cakes. The tridge $[\![B, C, E]\!]$ has the structure of Fig. 13d because $D$ is nested inside $C$, but $B$ is neither enclosed by, nor encloses, either $D$ or $C$ (Fig. 19e.) The structure of class $[\![B, C, D, F]\!]$ is not uniquely determined from Fig. 19e because $F$ could either be inside $D$ as shown or moved outside $D$ via a homotopic distortion (i.e. without crossing a red curve). However, as we will later see in Fig. 25a (and Fig. 26a), $D$ and $F$ are iterates of $C$ and $B$, respectively. Since $C$ and $B$ are not nested (Fig. 19e), $D$ and $F$ are not nested either. Further analysis of the nesting of $B$, $C$, $D$, and $F$ from Fig. 19e shows that $[\![B, C, D, F]\!]$ has the structure of Fig. 13h. Finally, the tridge $[\![A, E, E]\!]$ also needs extra consideration because the two $E$ circles could be nested or not. However, as we will later see in Fig. 26a, these two $E$ curves are iterates of $D$ and $F$, which we have just seen are not nested. Hence, the two $E$ curves are not nested. Since both of these $E$ curves are nested inside $A$, $[\![A, E, E]\!]$ has the structure of Fig. 13g.

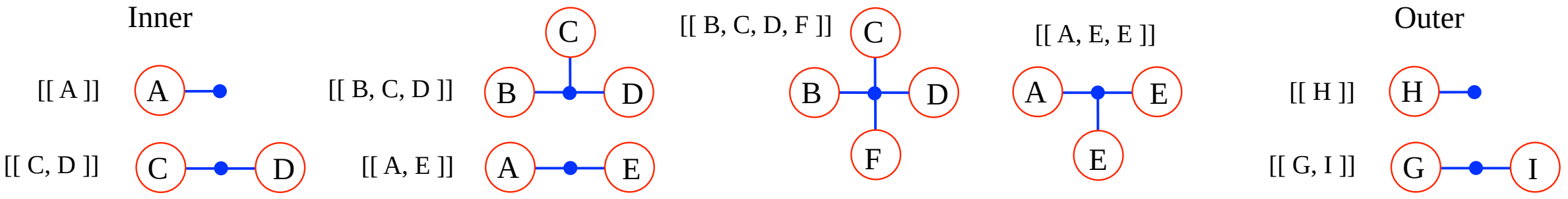

**Figure 20.** *Inner and outer bridge classes for Example 3.*

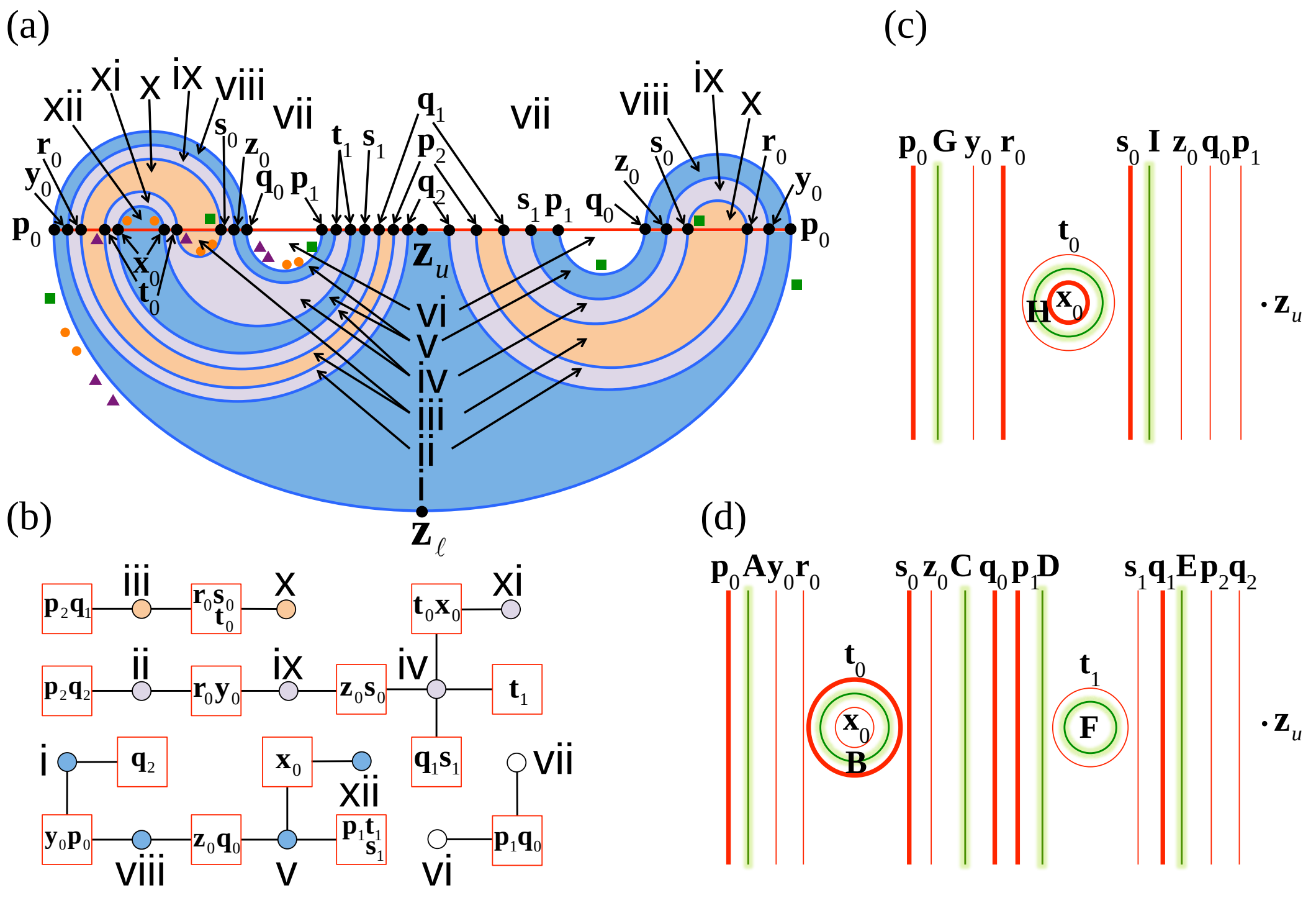

**Figure 21.** *Secondary division for Example 3. a) The division of phase space. b) The connection graph. c) The inner stable division. d) The outer stable division. In (c) and (d), the curves from the primary division are shown in bold. The boundary classes are shown in green.*

**3.1.5. Secondary division and bridge class iterates.** Analogous to 2D maps, we construct the secondary division of phase space by cutting along the surfaces:

1. the stable component of the trellis, i.e. the cap $T^S = W^S[\mathbf{p}_0]$;
2. the forward iterate of every bridge with a pseudoneighbor in its interior.

The secondary division is shown in cross-section in Fig. 21a. It has 12 regions. The connection graph between these regions is shown in Fig. 21b, where two regions are connected if they share a common boundary domain in the fundamental stable annulus. These common boundary domains are illustrated with red squares containing the boundary curves that define the domain in $T^S$.

We now map all active bridge classes forward one iterate. We begin with bridge class $[\![A]\!]$, as shown in Fig. 22. We can represent the bridge class $[\![A]\!]$ by a surface $\mathcal{S}$ with a single boundary circle $\mathcal{C}$ of class $A$. According to Fig. 19e, $\mathcal{C}$ can be chosen to lie within the stable domain $W^S[\mathbf{p}_0, \mathbf{s}_0, \mathbf{t}_0]$ of the primary division, and furthermore it can be chosen to be a small perturbation of $\mathbf{p}_0$. We represent curve $\mathcal{C}$ in Fig. 22a by the green circle around $\mathbf{p}_0$. The forward iterate $M(\mathcal{C})$ of $\mathcal{C}$ thus lies within the stable domain $W^S[\mathbf{p}_1, \mathbf{s}_1, \mathbf{t}_1]$ of the secondary division, illustrated by the red box in Fig. 22a enclosing $\mathbf{p}_1, \mathbf{s}_1, \mathbf{t}_1$. We further deduce that $M(\mathcal{C})$ can be represented by a perturbation of the circle $\mathbf{p}_1$, illustrated by the green circle around $\mathbf{p}_1$ in Fig. 22a. According to Fig. 21c, a perturbation of $\mathbf{p}_1$ into the region $W^S[\mathbf{p}_1, \mathbf{s}_1, \mathbf{t}_1]$ has boundary class $D$, which we place below the red box in Fig. 22a.

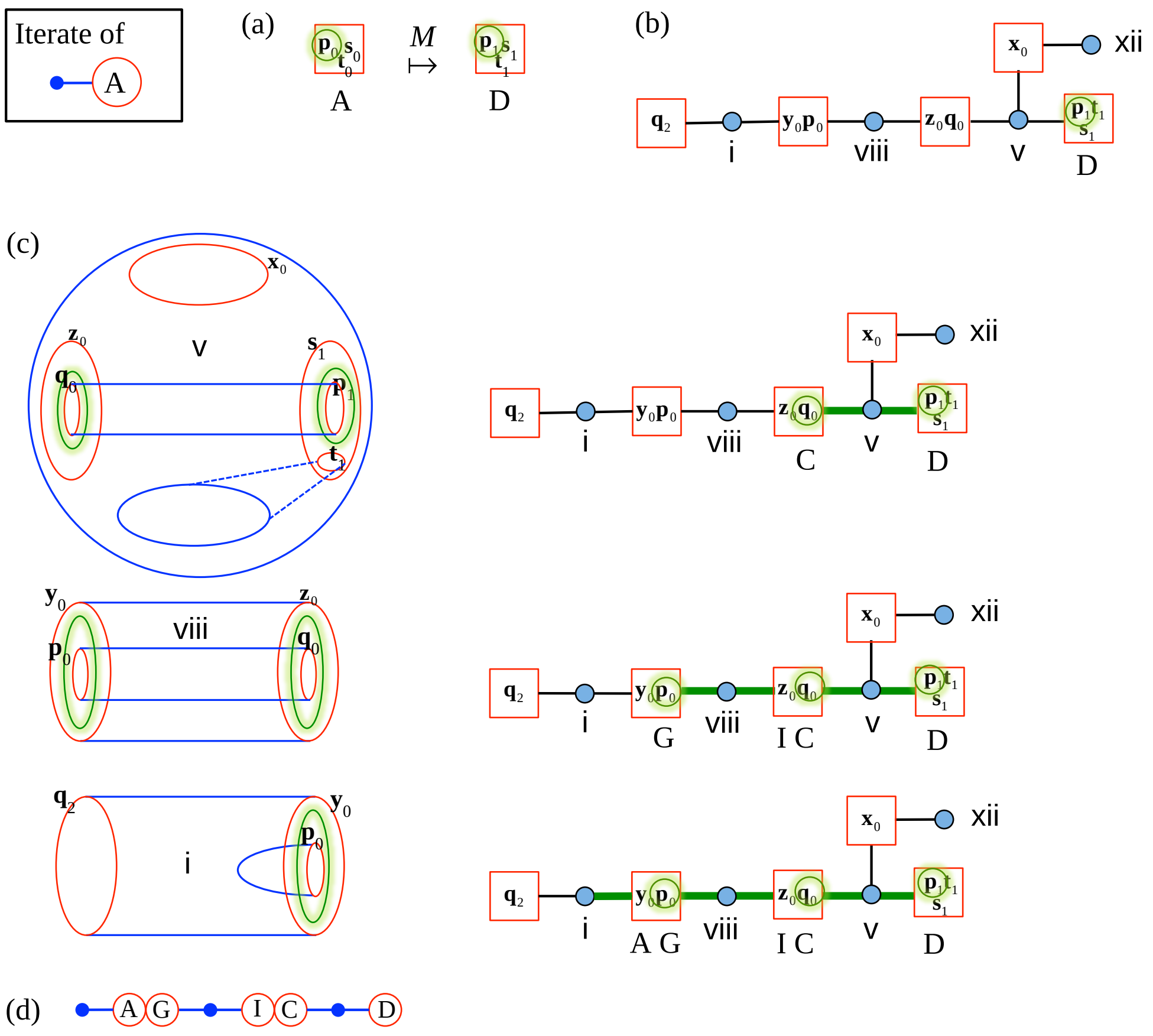

**Figure 22.** *Steps used to construct the iterate of class* $[\![A]\!]$.

We pick the component of the connection graph from Fig. 21b that includes $W^S[\mathbf{p}_1, \mathbf{s}_1, \mathbf{t}_1]$, and reproduce it in Fig. 22b, along with the green circle representing $M(\mathcal{C})$. The forward iterate of $\mathcal{S}$ must thus lie within the union of all the regions within this component. Recall in the 2D case that there were always two special squares (shaded green) in the connection graph and the corresponding bridge simply joined these squares by the shortest path within the connection graph. In 3D, it is not so simple, since there may in principle be any number of special squares, with green circles inside, and the manner in which they are joined is by no means obvious. What we do know in this example is that $M(\mathcal{S})$ must enter region v of the connection graph, but where it goes from there cannot be determined by the connection graph alone. We must examine the topology of region v in more detail; this is shown in the upper left of Fig. 22c. The boundary of region v is illustrated by the solid red and blue curves. The red curves delineate the three stable domains of the boundary (corresponding to the three red squares attached to vertex v in Fig. 22b), and the blue curves delineate the remaining two unstable domains of the boundary. The unstable domain $W^U[\mathbf{x}_0, \mathbf{z}_0, \mathbf{s}_1, \mathbf{t}_1]$ forms the outer sphere of the image, but part of this outer boundary is pushed inward, connecting to the circle $\mathbf{t}_1$. The second unstable domain forms a cylinder $W^U[\mathbf{q}_0, \mathbf{p}_1]$ passing through the interior of the sphere. The boundary of region v is thus a genus-two torus. It should be noted that the sketch of region v in Fig. 22c is not unique; for example, the "outer" part of the boundary could be formed by the unstable cylinder $W^U[\mathbf{q}_0, \mathbf{p}_1]$ rather than $W^U[\mathbf{x}_0, \mathbf{z}_0, \mathbf{s}_1, \mathbf{t}_1]$.

In region v in Fig. 22c, $M(\mathcal{C})$ is represented by the green circle enclosing $\mathbf{p}_1$. Since $M(\mathcal{S})$ does not intersect the cylinder $W^U[\mathbf{q}_0, \mathbf{p}_1]$, $M(\mathcal{S})$ must enclose the cylinder, forming a circular intersection with the stable domain $W^S[\mathbf{q}_0, \mathbf{z}_0]$; this intersection is shown by the green circle enclosing $\mathbf{q}_0$. Thus, there is a piece of $M(\mathcal{S})$ that forms a cylinder connecting the two green circles in region v of Fig. 22c. (To avoid clutter, this cylinder is not shown.) In principle, $M(\mathcal{S})$ could form a more complicated surface within region v, for example, one that reaches up and intersects the other stable domain $W^S[\mathbf{x}_0]$. However, this is not *forced* to occur. Throughout this process, we seek the simplest possible topology of $M(\mathcal{S})$ and avoid unnecessary intersections. The joining of the two green circles by $M(\mathcal{S})$ is now shown as the thick green segment within the connection graph, connecting the circle around $\mathbf{p}_1$ to the circle around $\mathbf{q}_0$. Next, Fig. 21c shows that the green circle around $\mathbf{q}_0$ has boundary class $C$. (We use the inner secondary division because region v is an inner region, as seen in Fig. 21a.) We thus place a $C$ below the green circle enclosing $\mathbf{q}_0$ in the connection graph.

Next, $M(\mathcal{S})$ must enter region viii, whose topology is shown on the next line in Fig. 22c. The green circle enclosing $\mathbf{q}_0$ in region v is now reproduced as the green circle enclosing $\mathbf{q}_0$ on the right of region viii. $M(\mathcal{S})$ must then join the right green circle to the left green circle enclosing $\mathbf{p}_0$. This is again recorded by a green segment in the connectivity graph. The two green circles enclosing $\mathbf{q}_0$ and $\mathbf{p}_0$ have boundary classes $I$ and $G$ when viewed from region viii (an outer region), as can be verified in Fig. 21d. We thus place $I$ and $G$ below the appropriate green circles in the connection graph.

Finally, $M(\mathcal{S})$ enters region i, shown on the next line. The green circle enclosing $\mathbf{p}_0$ in region viii is now reproduced in region i. Due to the topology of region i, the piece of $M(\mathcal{S})$ contained in region i is not forced to intersect the boundary of region i anywhere else. Rather $M(\mathcal{S})$ in region i can form a cap, with the green circle as its single boundary component. This cap wraps around the unstable domain $W^U[\mathbf{p}_0]$. This is represented in the connection graph by the green line segment terminating at the dot of region i. The green circle around $\mathbf{p}_0$ has boundary class $A$ when viewed from region i, an inner region. This completes the construction of $M(\mathcal{S})$.

The homotopy class of $M(\mathcal{S})$ can now be read off from the connection graph. Each green segment within a particular region becomes a bridge class with boundary classes shown below the corresponding green circles. Translating this into our typical notation for bridge classes, we find the homotopy class shown in Fig. 22d. Analogous to the 2D dynamics, two adjacent red circles represent the concatenation of two bridge classes at their corresponding boundary circles. Note that there is exactly one free boundary circle, which is not concatenated with anything else. This is as it must be, since the original bridge class $[\![A]\!]$ itself has a single boundary circle, and $M([\![A]\!])$ has the same number of boundary circles as $[\![A]\!]$ itself.

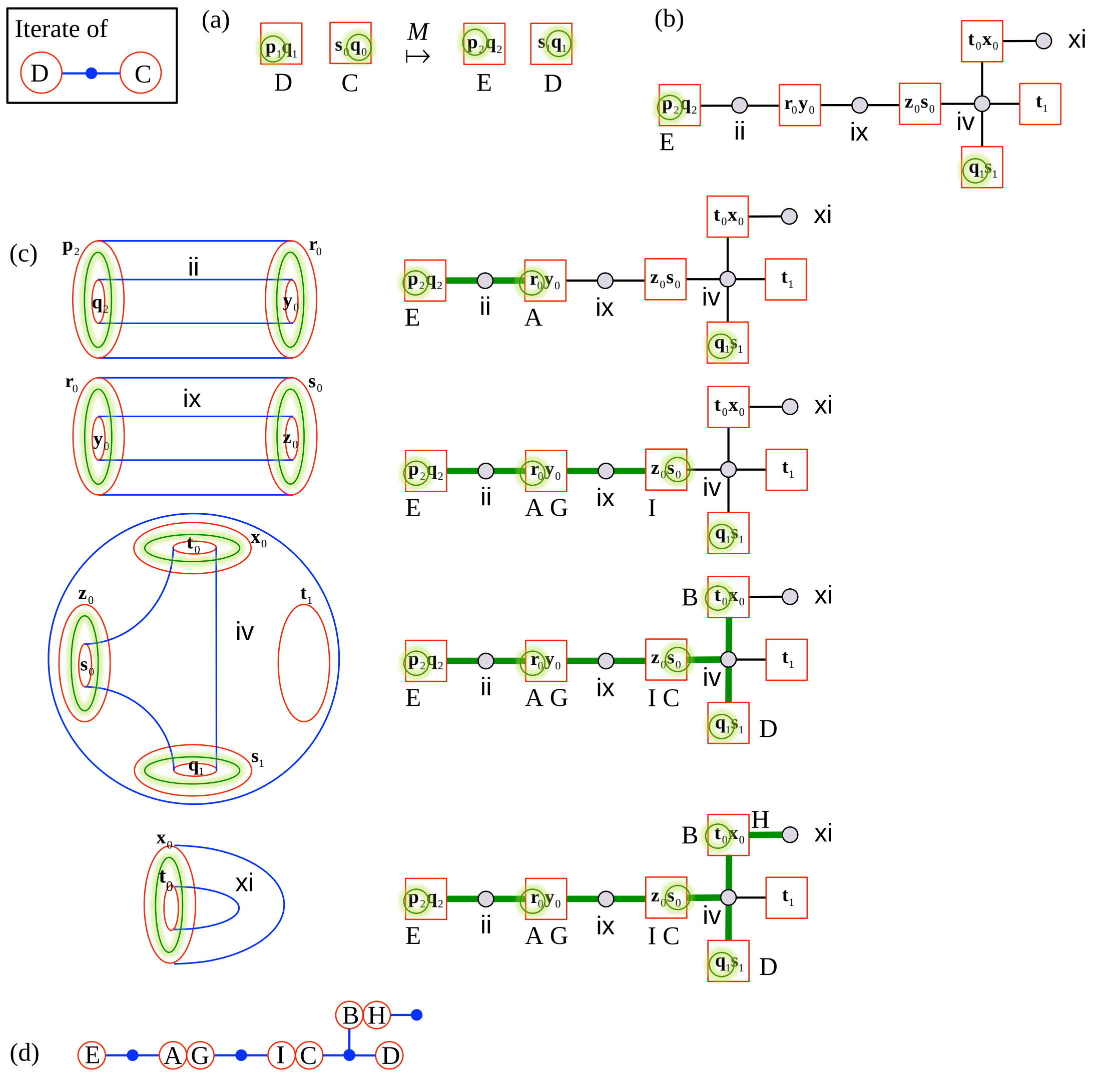

**Figure 23.** *Steps used to construct the iterate of class* $[\![C, D]\!]$.

We next iterate the active class $[\![C, D]\!]$ forward. See Fig. 23. The boundary classes $C$ and $D$ lie in the stable domains $W^S[\mathbf{s}_0, \mathbf{q}_0]$ and $W^S[\mathbf{p}_1, \mathbf{q}_1]$, respectively, within the primary division (Fig. 19e). Let $\mathcal{S}$ be a surface representing $[\![C, D]\!]$. Then the forward iterate $M(\mathcal{S})$ will have boundary circles in $W^S[\mathbf{s}_1, \mathbf{q}_1]$ and $W^S[\mathbf{p}_2, \mathbf{q}_2]$ within the secondary division (Fig. 23a). The boundary circles are again illustrated in green. We must then join these boundary circles within the connection graph shown in Fig. 23b. The shortest path, appropriate for the 2D case, is not the correct way to join these two circles. Fig. 23c shows the correct way to construct the subgraph joining these circles, following the same technique as in Fig. 22c. This process is trivial until region iv in the connection graph (third row of Fig. 23c). In region iv, we must join the green circle enclosing $\mathbf{s}_0$ to the green circle enclosing $\mathbf{q}_1$, using a 2D surface that does not intersect the unstable boundary of region iv. Due to the presence of $W^U[\mathbf{s}_0, \mathbf{q}_1, \mathbf{t}_0]$, joining these two green circles requires that the surface also intersect $W^S[\mathbf{x}_0, \mathbf{t}_0]$ in the green circle shown in Fig. 23c. This is the first example where the topology of the region forces a branching in the representation of $M(\mathcal{S})$. To complete the construction, $M(\mathcal{S})$ in region xi must form a cap with boundary circle enclosing $\mathbf{t}_0$. Below each green circle in the connection graphs in Fig. 23c, we place the corresponding boundary classes, viewed from both the inner and outer zones as appropriate. From this information, $M([\![C, D]\!])$ is summarized in Fig. 23d.

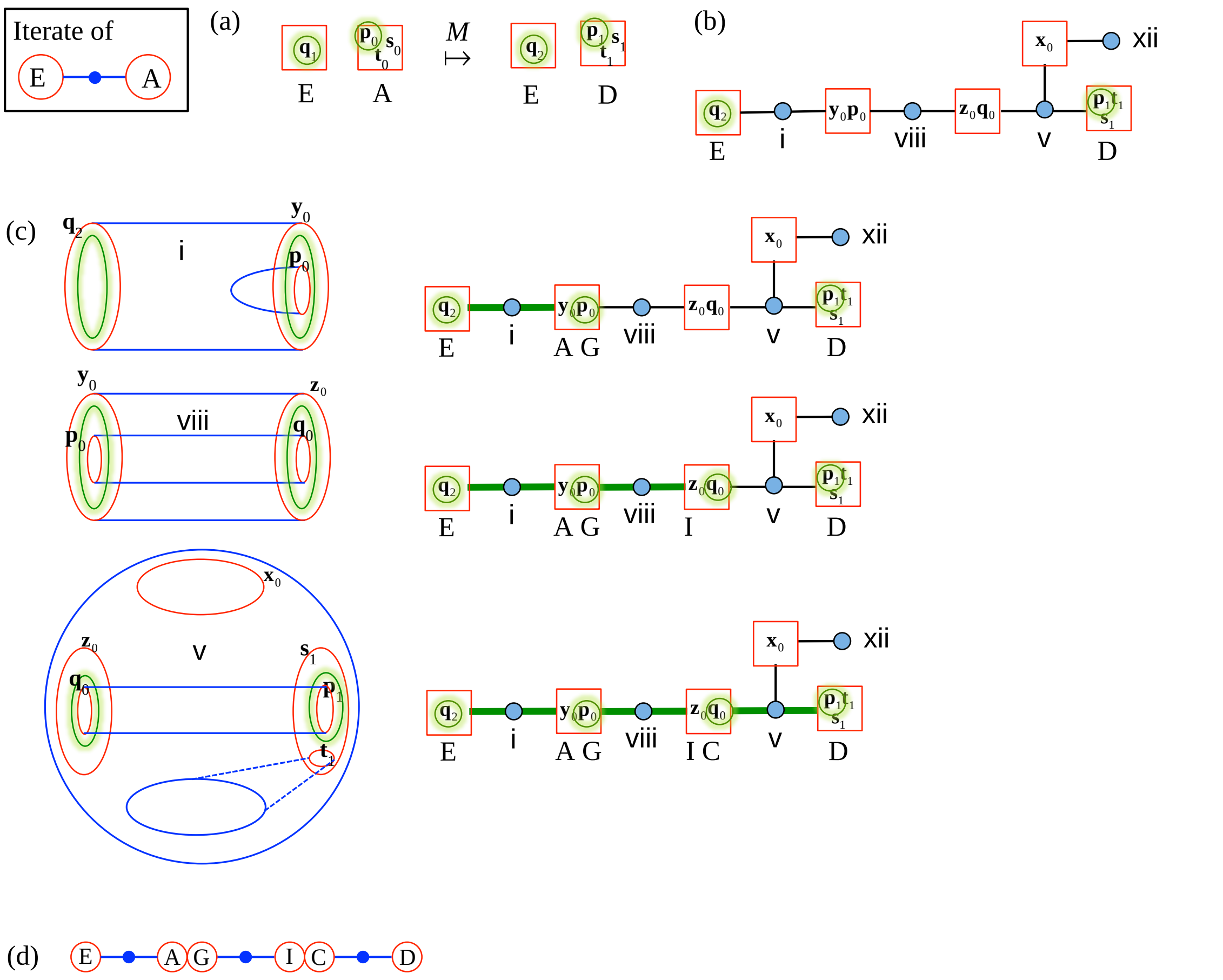

**Figure 24.** *Steps used to construct the iterate of class* $[\![A, E]\!]$.

Fig. 24 shows how to construct the iterate of $[\![A, E]\!]$. This case is almost identical to that of $[\![A]\!]$ in Fig. 22, except for the presence of an additional boundary class.

Next, Fig. 25 shows how to construct the iterate of $[\![B, C, D]\!]$. This case is almost identical to that of $[\![C, D]\!]$ in Fig. 23, except for the additional boundary class $B$. This boundary class maps forward to boundary class $F$ in domain $W^S[\mathbf{t}_1]$ (Fig. 25a). Consequently, in region iv (third line of Fig. 23c), the green boundary circle in $W^S[\mathbf{t}_1]$ is connected to the green subgraph. This results in the inclusion of the boundary class $F$ on the right of Fig. 25d, the only distinction from Fig. 23d. Thus, the initial bridge class and its iterate both have three free boundary classes in Fig. 25. Note that Fig. 25d contains the bridge class $[\![B, C, D, F]\!]$, with four boundary classes. We iterate this class next.

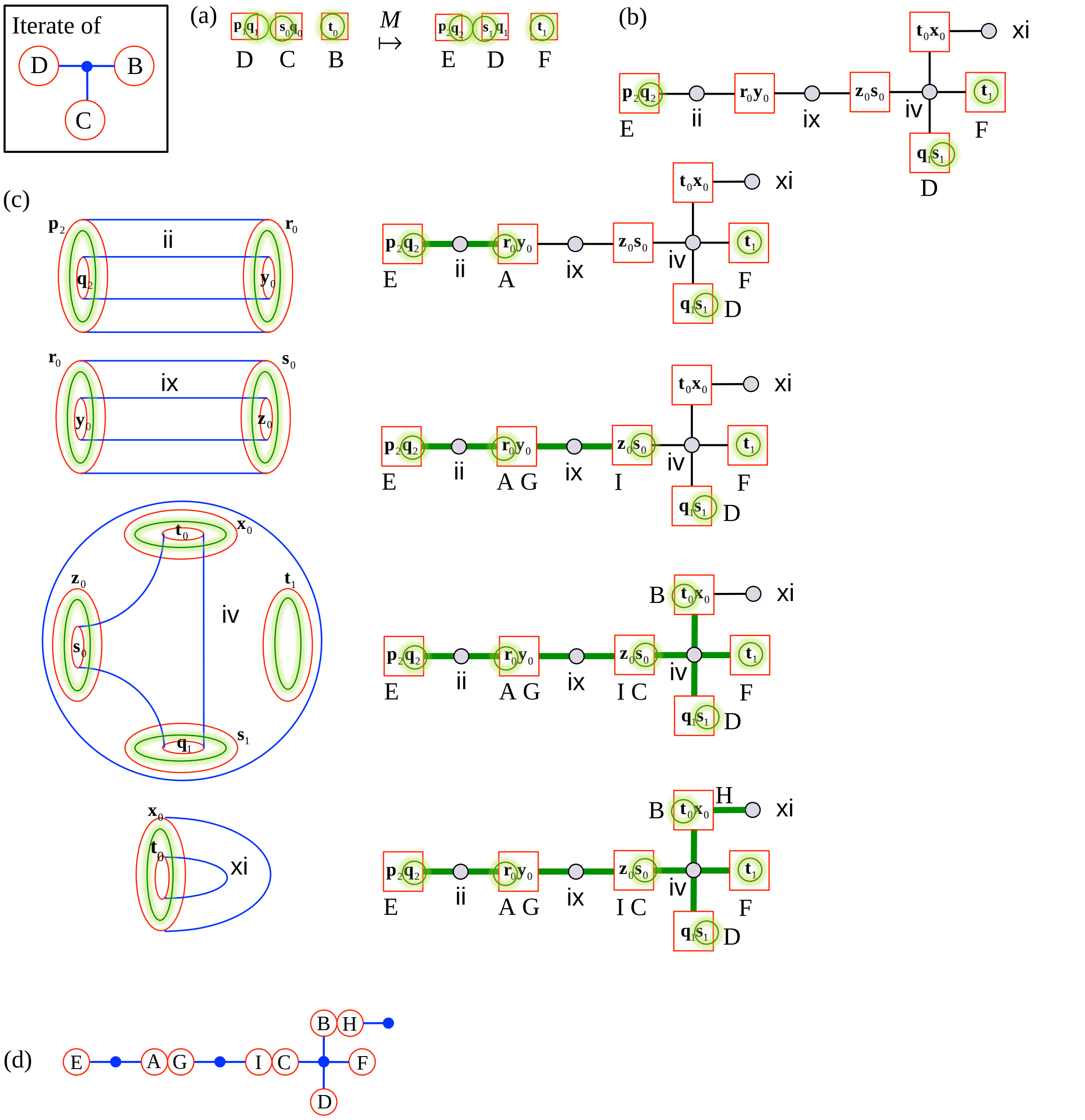

**Figure 25.** *Steps used to construct the iterate of* $[\![B, C, D]\!]$.

Fig. 26 shows how to construct the iterate of $[\![B, C, D, F]\!]$. In Fig. 26a, the circle representing boundary class $D$ is a perturbation of $\mathbf{p}_1$ (Fig. 19e), and it maps forward to a circle that is a perturbation of $\mathbf{p}_2$ in the secondary division. This circle is of boundary class $E$

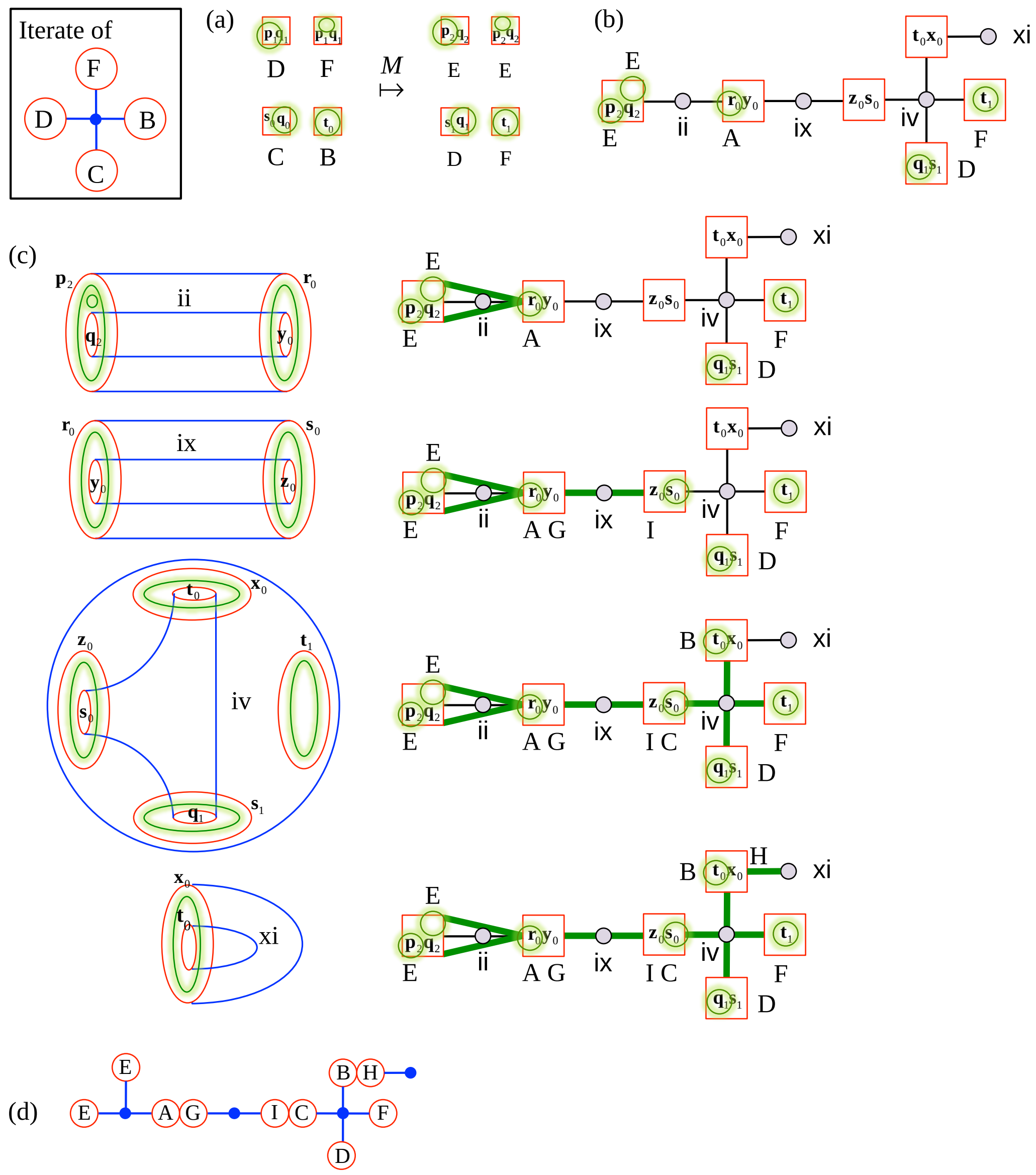

**Figure 26.** *Steps used to construct the iterate of* $[\![B, C, D, F]\!]$.

(Fig. 21c). On the other hand, the circle representing boundary class $F$ is not a perturbation of any circle from the primary division (Fig. 19e), but rather is contractible in the domain $W^S[\mathbf{p}_1, \mathbf{q}_1]$. Thus it maps forward to a contractible circle in the domain $W^S[\mathbf{p}_2, \mathbf{q}_2]$ of the secondary division. This circle has boundary class E (Fig. 21c.) Thus, there are two boundary circles in the forward iterate of $[\![B, C, D, F]\!]$ which have the same boundary class, $E$. This is illustrated in Fig. 26b by the two green circles in the leftmost square, one enclosing $\mathbf{p}_2$ and one enclosing nothing. Fig. 26c shows these two green boundary circles on the left of region

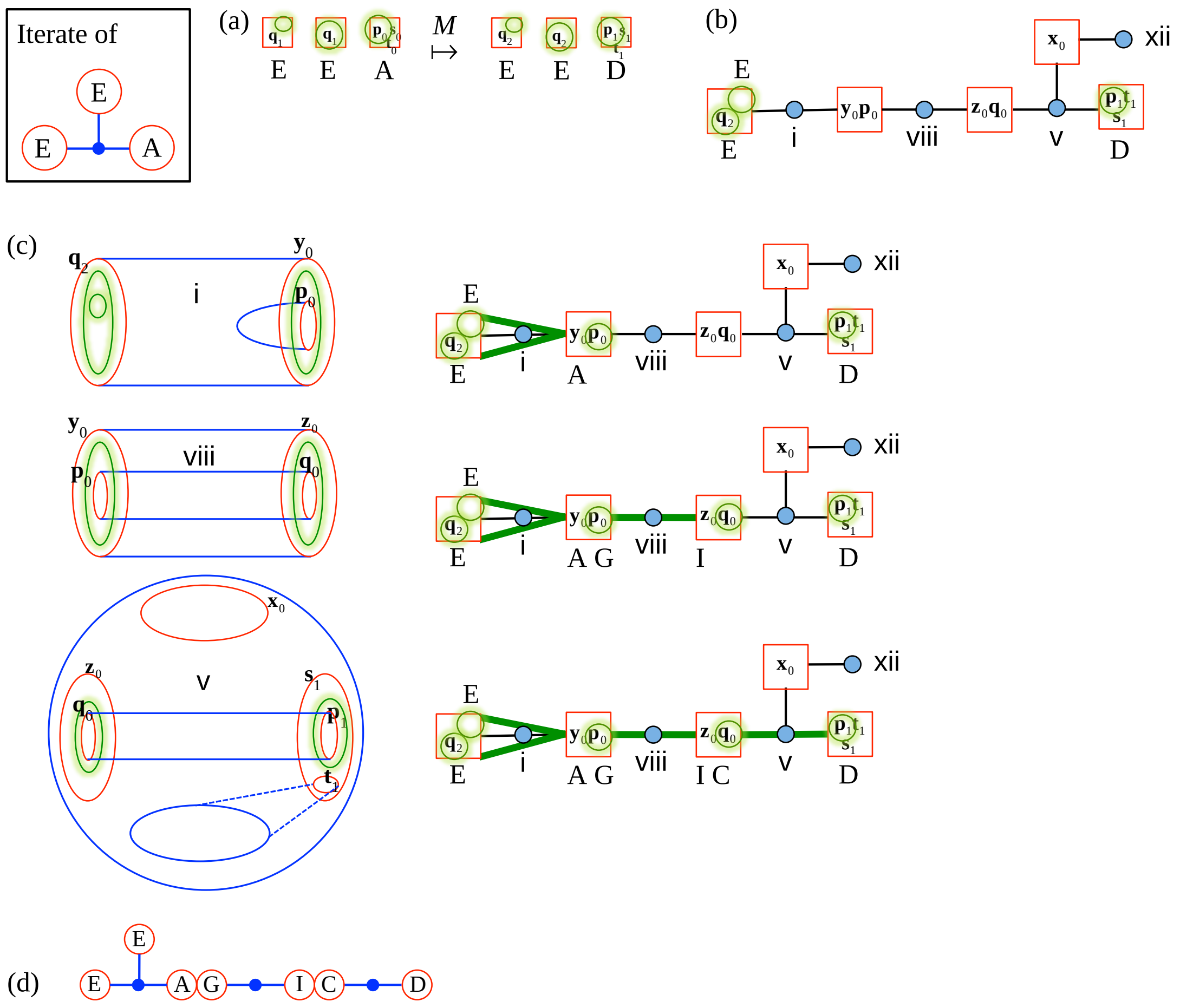

**Figure 27.** *Steps used to construct the iterate of class* $[\![A, E, E]\!]$.

ii. The manifold $M(\mathcal{S})$ then joins these two boundary circles to the boundary circle on the right of region ii. This is illustrated by the double green line in the connection graph. Note that the additional boundary circle on the left does not force any additional intersections of $M\mathcal{S}$ with $T^S$. From this point on, the construction is identical to that in Fig. 25. The final iterate, shown in Fig. 26d, has four free boundary classes. It is identical to Fig. 25d with an additional free boundary class $E$ attached to the leftmost vertex.

Finally, Fig. 27 shows how to construct the iterate of $[\![E, E, A]\!]$. This is identical to the iterate of $[\![E, A]\!]$ in Fig. 24, except for the additional boundary class $E$, which essentially acts as a spectator.

### 3.1.6. Transition matrix and topological entropy.

The complete dynamics for all active bridge classes is summarized in Fig. 28. From this, the transition matrix for just the active

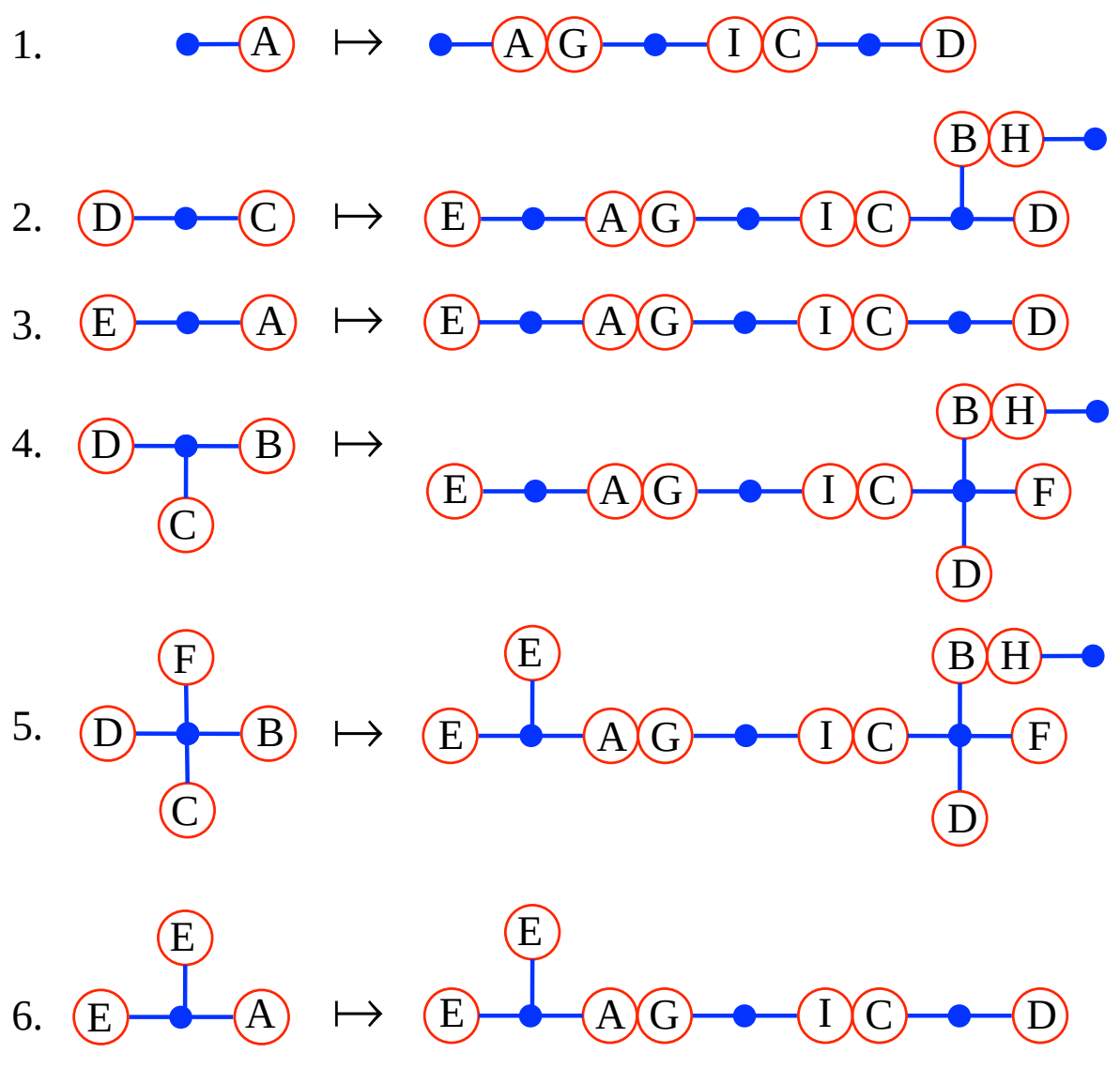

**Figure 28.** *Active bridge class dynamics for Example 3.*

classes is

$$T = \begin{pmatrix} 1 & 0 & 0 & 0 & 0 & 0 \\ 1 & 0 & 1 & 0 & 0 & 1 \\ 0 & 1 & 1 & 1 & 0 & 0 \\ 0 & 1 & 0 & 0 & 0 & 0 \\ 0 & 0 & 0 & 1 & 1 & 0 \\ 0 & 0 & 0 & 0 & 1 & 1 \end{pmatrix}, \tag{12}$$

where the ordering of classes along the columns and rows is given by the numbering in Fig. 28. As seen in this matrix, every active classes generates two active classes, yielding the topological entropy

$$h_{\text{top}} = \ln 2. \tag{13}$$

This result may seem surprising. Recall that the 2D cross-section (Fig. 16a) of the Example 3 trellis resembles Fig. 3a on the left ($h_{\text{top}} = \ln 2.2695$) and Fig. 2a on the right ($h_{\text{top}} = \ln 2$). One might therefore expect the topological entropy of Example 3 to be at least $\ln 2.2695$, the maximum of the 2D topological entropies. Indeed, this would be the case if the cross-sectional plane shown in Fig. 16a were invariant under the dynamics. However, Example 3 does not assume that this plane is invariant. Relaxing this constraint means that some heteroclinic intersections that would have been topologically forced in the plane are no longer forced, because some 2D bridges, or parts of 2D bridges, are free to leave the plane under homotopic distortion, reducing the number of heteroclinic intersection curves. This point is illustrated more clearly in the following simpler example.

### 3.2. Example 4: A zero topological entropy case .

We consider a simpler tangle than those considered previously to illustrate how we can lose topological entropy entirely when we extend a tangle in 2D into 3D.

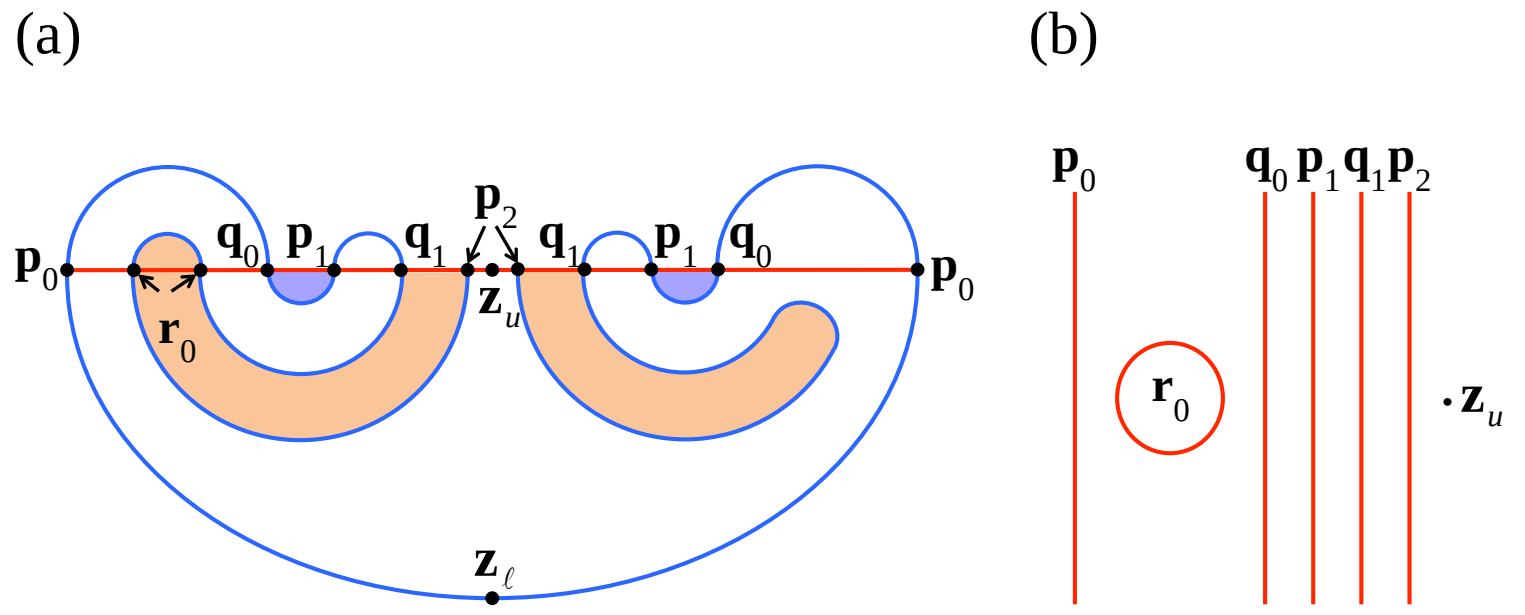

**Figure 29.** *The trellis for Example 4. Shown are (a) the cross-section and (b) the top-down view of the stable cap.*

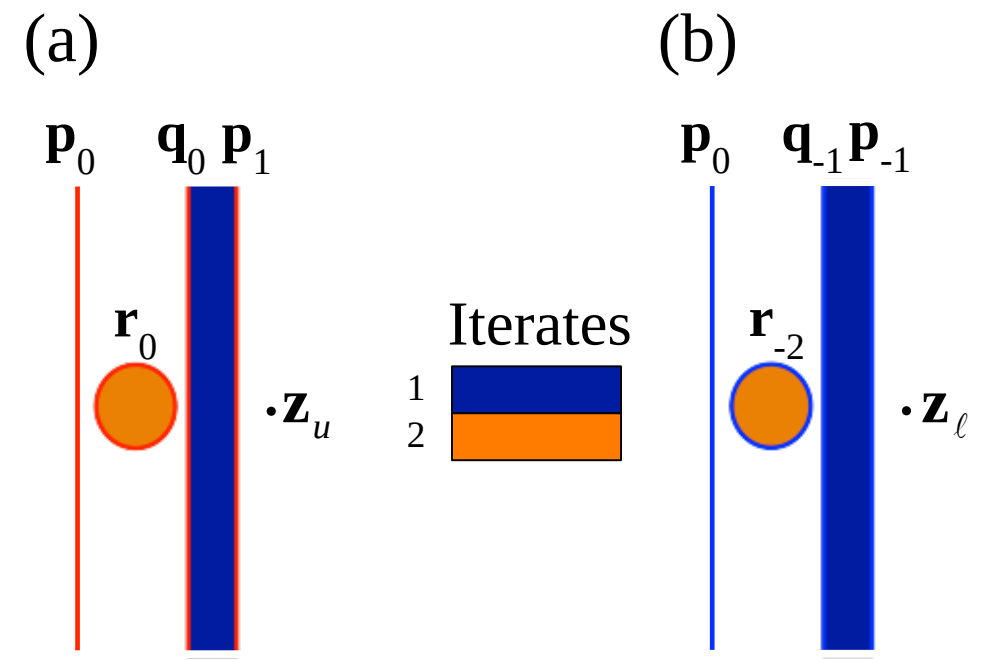

**Figure 30.** *(a) Backward and (b) forward escape-time plots for Example 4.*

**3.2.1. Trellis.** Consider the asymmetric trellis whose cross section is shown in Fig. 29a having a complete horseshoe on the left of the $z$ axis and an incomplete horseshoe on the right of the $z$ axis. A top-down view of the stable cap, with heteroclinic intersections, is shown in Fig. 29b.

**3.2.2. Escape-time plots and pseudoneighbors.** From the backward escape-time plot in Fig. 30a and the forward escape-time plot in Fig. 30b, we see that there are three pairs of pseudoneighbors: (i) $\mathbf{q}_n$ and $\mathbf{p}_n$, (ii) $\mathbf{q}_n$ and $\mathbf{p}_{n+1}$, (iii) $\mathbf{r}_n$ with itself.

**3.2.3. Holes.** With pseudoneighbors identified, we create three families of holes, as shown in Fig. 31a: (i) One family of holes (green squares) is a perturbation of $\mathbf{p}_n$, (ii) one (purple triangles) is a perturbation of $\mathbf{q}_n$, and (iii) one (orange circles) is a perturbation of $\mathbf{r}_n$.

**3.2.4. Primary division and bridge classes.** Figure 31b shows the primary division of phase space. Figures 31c and 31d show the corresponding inner and outer divisions of the stable cap, respectively. The green circles in Figs. 31c and 31d are the boundary classes for this example. The corresponding inner and outer bridge-classes, connecting the boundary classes, are shown in Fig. 31e.

**3.2.5. Secondary division and bridge class iterates.** Iterating forward every active bridge in the primary division (i.e. every bridge with a pseudoneighbor in its interior), and using these to cut phase space, we obtain the secondary division (Fig. 32a). The corresponding connection graph is shown in Fig. 32b. Figures 32c and 32d show the inner and outer secondary divisions of the stable cap, along with the boundary classes, shown in green. Using the information that we've obtained thus far, one can compute the bridge-class iterates, as in Sect. 3.1. Fig. 32e shows the iterates of all four inner bridge-classes from Fig. 31e. Of these, the top two are active and the bottom two are inert. (All outer bridge-classes from Fig. 31e are inert.)

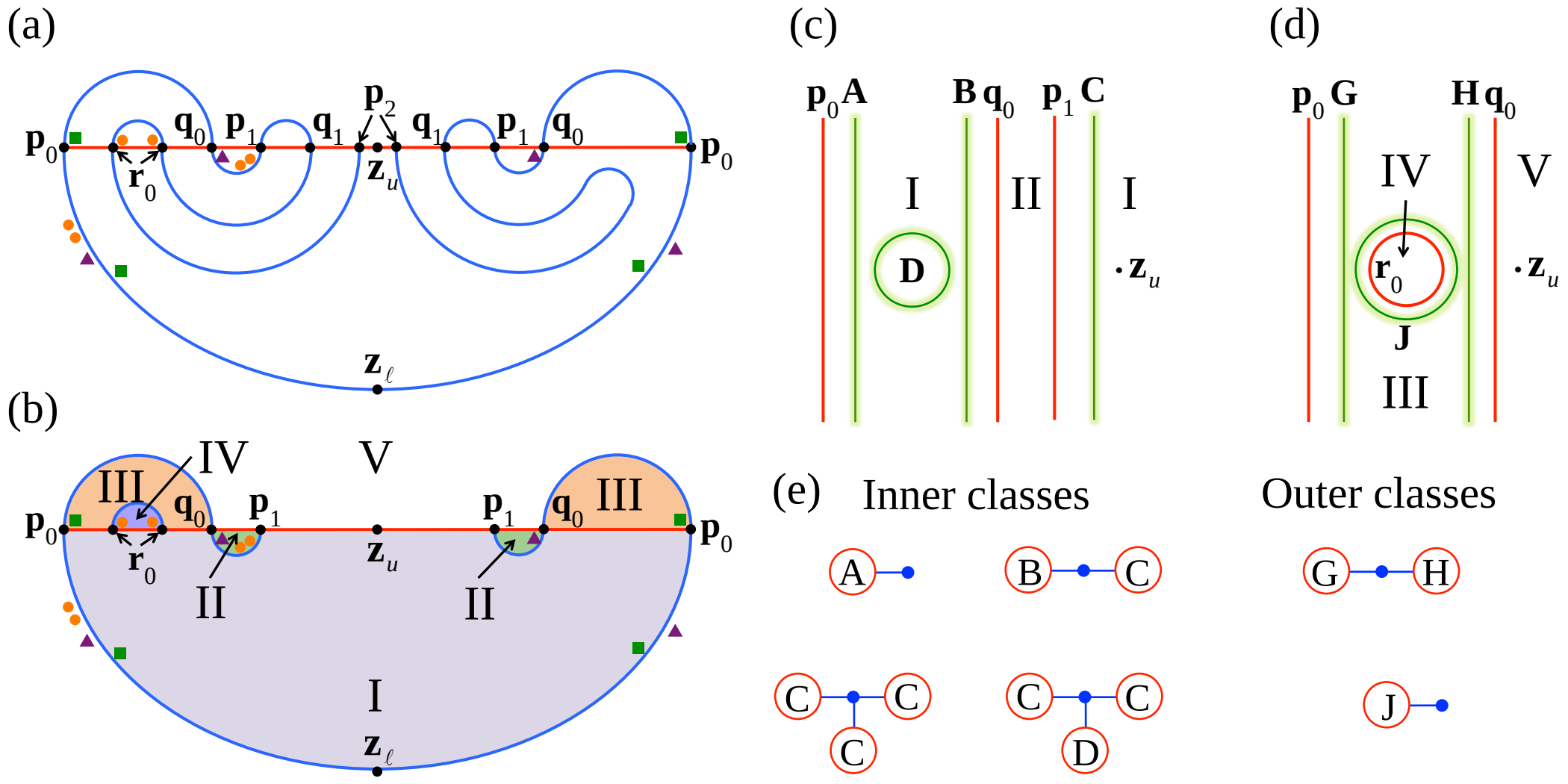

**Figure 31.** *a) Trellis for Example 4 with holes. b) The primary division of phase space. c) The primary inner stable division. d) The primary outer stable division. e) The inner and outer bridge classes.*

**3.2.6. Transition matrix and topological entropy.** The transition matrix for all inner classes is

$$T = \begin{pmatrix} 1 & 0 & 0 & 0 \\ 1 & 0 & 0 & 0 \\ 0 & 1 & 0 & 0 \\ 0 & 0 & 1 & 1 \end{pmatrix}, \tag{14}$$

where the ordering of classes along the columns and rows is given by the numbering in Fig. 32e. The topological entropy is easily seen to be

$$h_{\text{top}} = \ln 1 = 0. \tag{15}$$

Thus, we obtain zero topological entropy for the 3D map whose trellis is shown in cross-section in Fig. 29a, even though the 2D trellis in Fig. 29a would yield a topological entropy of $\ln 2$ if the dynamics were confined to the plane. The fundamental reason the entropy is eliminated in 3D is as follows. The bridge $W^U[\mathbf{r}_0]$ that encloses region iv in Fig. 31a is a cap. A lobe forced to pass through region ii by the 2D horseshoe dynamics, would form a macaroni-shaped bridge in 3D, wrapping around the cap $W^U[\mathbf{r}_0]$. This macaroni-shaped bridge could be homotopically pushed to the side, around the cap, and then pressed through the stable manifold, thereby eliminating the heteroclinic intersections forming the macaroni-shaped bridge itself. In this manner, all of the subsequent heteroclinic intersections forced by the 2D horseshoe in Fig. 31a can be eliminated by accessing the third dimension.

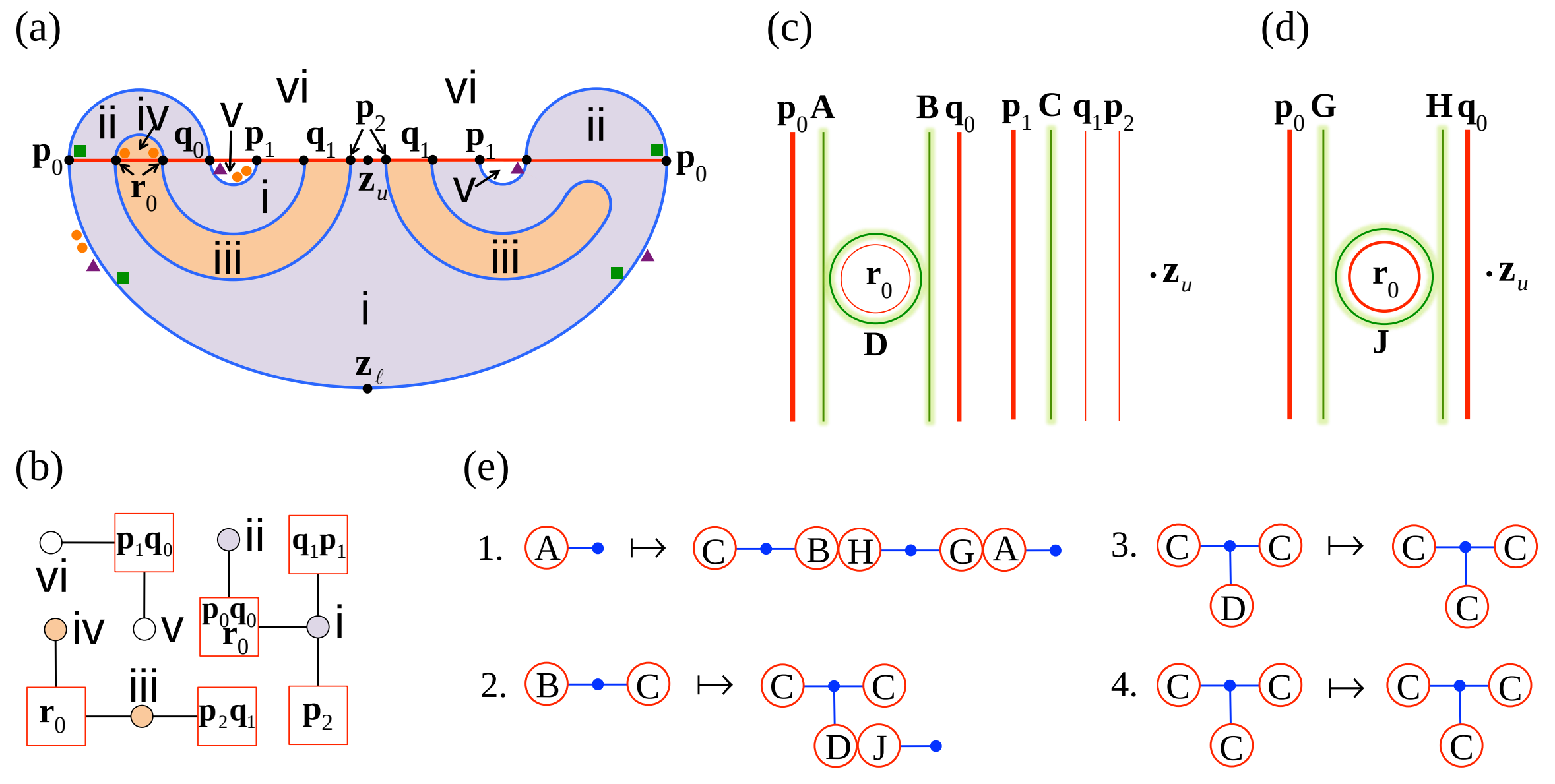

**Figure 32.** *a) The secondary division of phase space for Example 4. b) The connection graph. c) The secondary inner stable division. d) The secondary outer stable division. Curves from the primary division are shown in bold. Boundary classes are shown in green. e) Inner bridge class dynamics for Example 4.*

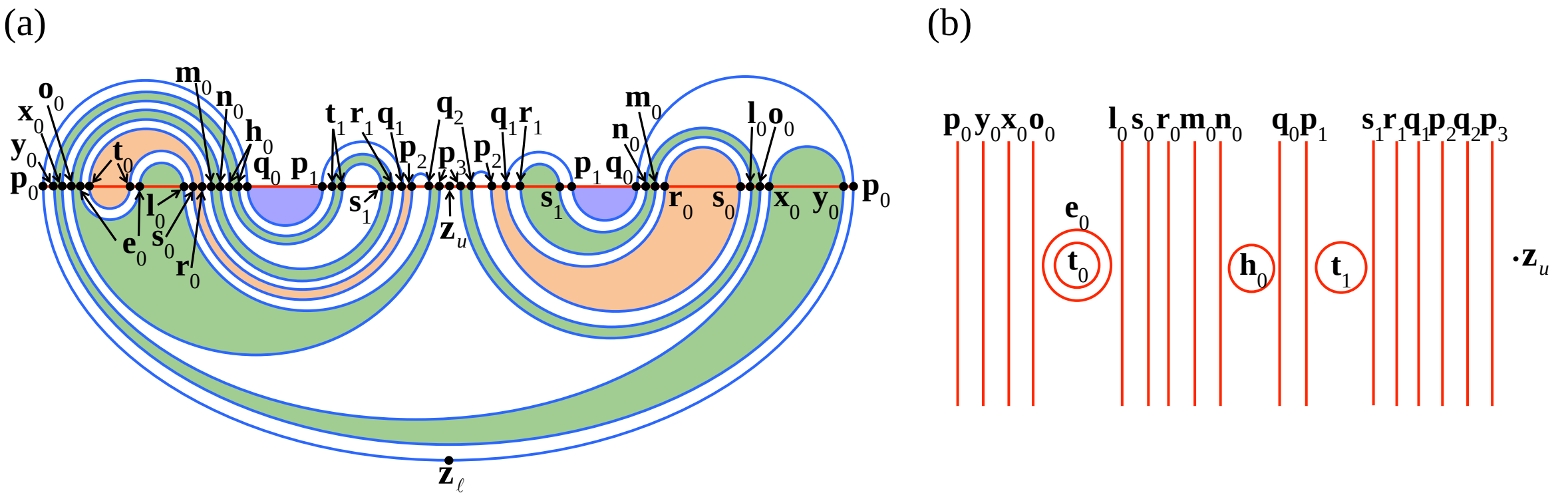

**Figure 33.** *The trellis for Example 5. Shown are (a) the cross-section and (b) the top-down view of the stable cap.*

**3.3. Example 5: An unambiguously 3D case.** With our methodology established, we turn our attention to a final trellis. It not only has no axis of symmetry, but it exhibits symbolic dynamics that is truly three-dimensional, dynamics which cannot be accommodated within a 2D model. We shall make this notion precise through an analysis of 1D versus 2D stretching rates in Sect. 4.

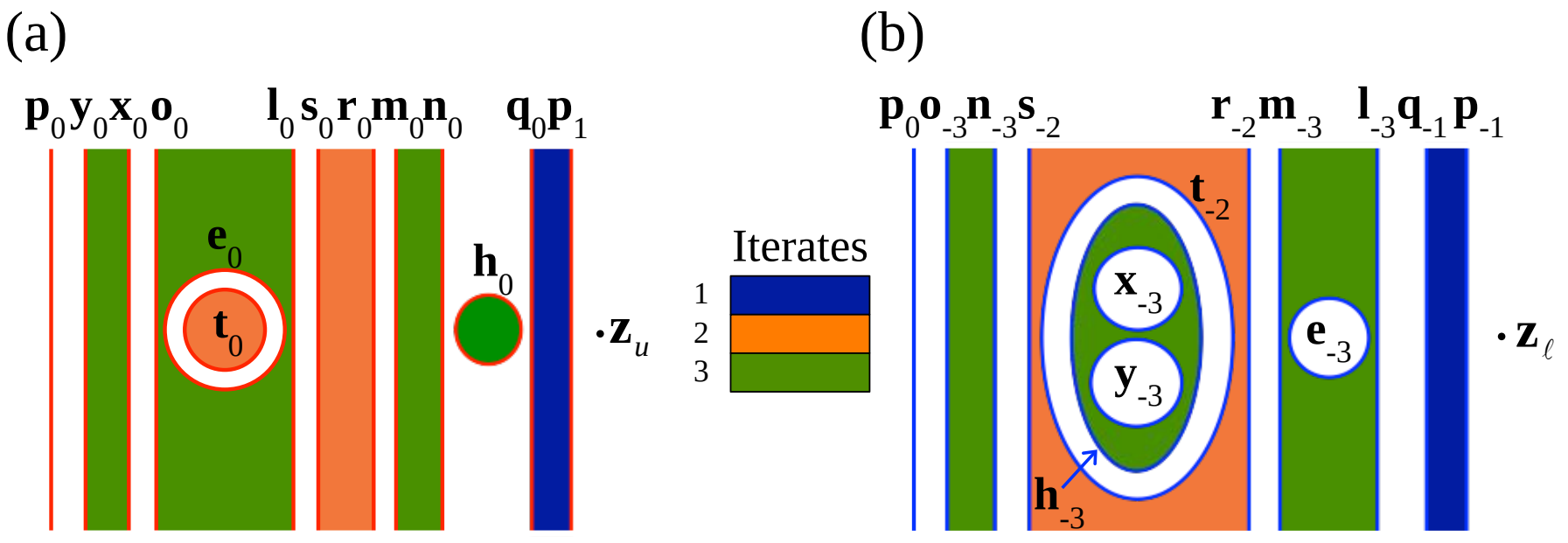

**Figure 34.** *(a) Backward and (b) forward escape-time plots for Example 5.*

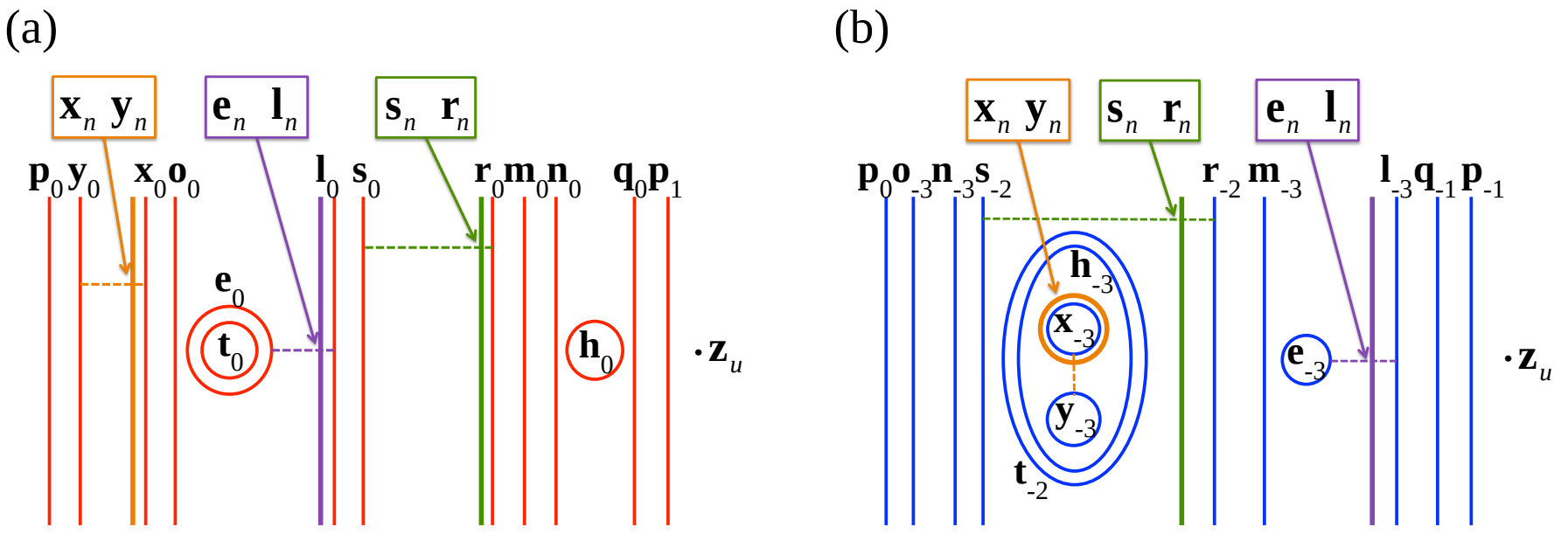

**Figure 35.** *Psuedoneighbor pairs and holes shown in the (a) stable and (b) unstable fundamental annuli.*

**3.3.1. Trellis.** Figures 33a and 33b show the trellis, via a cross-section and a top-down view as before. An important aspect of this trellis is that there is a second bridge that parallels the unstable cap $W^U[\mathbf{p}_0]$, containing the lower fixed point $\mathbf{z}_\ell$. This feature was not present in any of the prior trellises.

**3.3.2. Escape-time plots and pseudoneighbors.** Figures 34a and 34b show the forward and backward escape-time plots, which can be determined entirely from the data in Fig. 34. As before, the escape-time plots allow us to determine pseudoneighbors. In this example, there are three pseudoneighbor pairs, as shown in Fig. 35: (i) $\mathbf{s}_n$ and $\mathbf{r}_n$, (ii) $\mathbf{e}_n$ and $\boldsymbol{\ell}_n$, and (iii) $\mathbf{x}_n$ and $\mathbf{y}_n$.

**3.3.3. Holes.** With the pseudoneighbors located, we place holes next to them as shown in the fundamental annuli (Fig. 35) and in cross-section (Fig. 36a).

**3.3.4. Primary division and bridge classes.** Figure 36b shows the primary division of phase space. Similarly, Figs. 36c and 36f show the inner and outer stable primary divisions, respectively. Figs. 36d and 36g show the inner and outer boundary classes, respectively. Figs. 36e and 36h show the inner and outer bridge classes, respectively.

**3.3.5. Secondary division and bridge class iterates.** We iterate forward all active bridges in the primary division, and use these to construct the secondary division (Fig. 37a). Figure 37b shows the connection graph, in which regions of the division are connected if they share a boundary along the stable fundamental annulus. The inner and outer stable secondary divisions are shown in Figs. 37c and 37d, respectively. With this information, we construct the iterates of all of active bridge classes, which are shown in Fig. 37e.

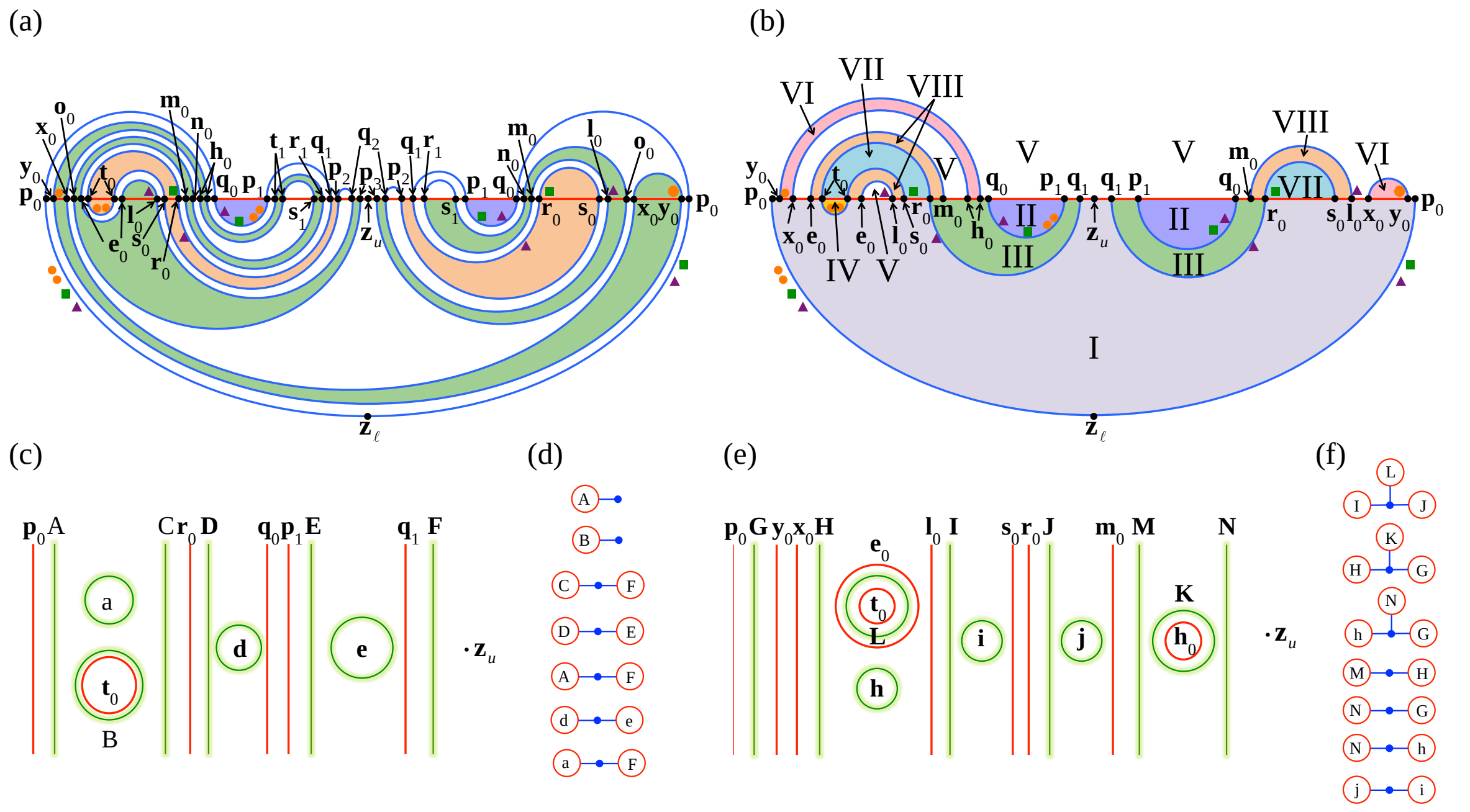

**Figure 36.** *a) Trellis for Example 5 with holes. b) The primary division of phase space. c) The primary inner stable division. d) Inner bridge classes. e) The primary outer stable division. f) Outer bridge classes.*

**3.3.6. Transition matrix and topological entropy.** The transition matrix for the active bridge classes is

$$T = \begin{pmatrix} 1 & 2 & 0 & 0 & 1 & 0 & 0 \\ 0 & 0 & 0 & 1 & 0 & 0 & 0 \\ 0 & 0 & 0 & 1 & 0 & 0 & 0 \\ 1 & 0 & 1 & 1 & 1 & 0 & 0 \\ 0 & 0 & 1 & 0 & 0 & 0 & 0 \\ 0 & 1 & 0 & 0 & 0 & 1 & 1 \\ 0 & 0 & 0 & 0 & 1 & 1 & 1 \end{pmatrix}, \tag{16}$$

with topological entropy

$$h_{\text{top}} = \ln 2.3593. \tag{17}$$

Notice that this is now greater than the topological entropy $\ln 2$ of the simple horseshoe.

**4. Symbolic analysis of 1D curves.** Thus far, we have based our analysis of 3D maps on the homotopy theory of surfaces, deriving symbolic dynamics for mapping surfaces forward in 3D. However, we can also analyse 3D maps using the homotopy theory of curves, i.e. using the usual fundamental group. In this approach, we use the same holes as before, i.e. ring-shaped holes punched near pseudoneighbor intersections. We then consider the homotopy classes of curves that begin and end on the stable cap $T^S$ and which do not pass through the holes.

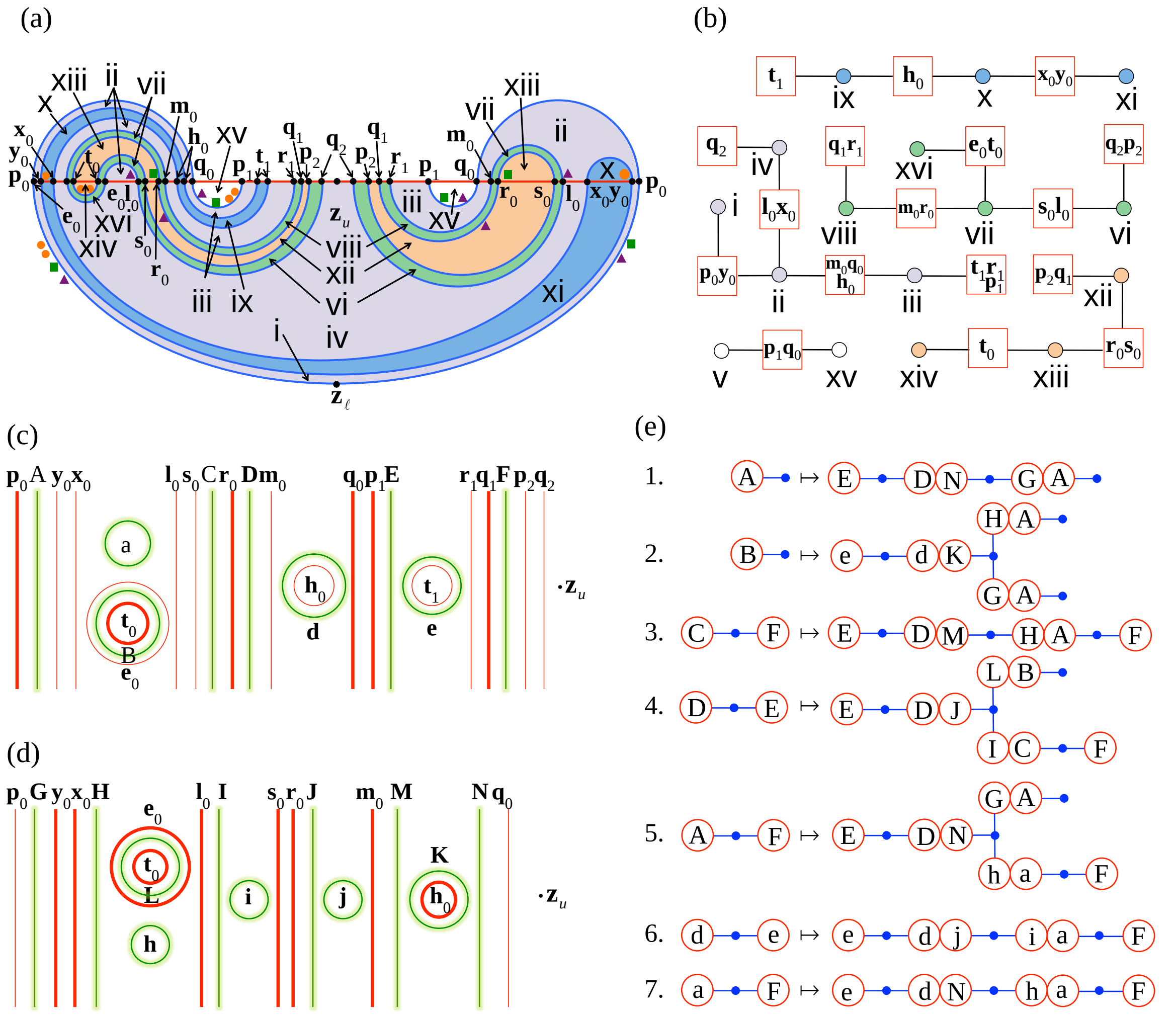

**Figure 37.** *a) The secondary division of phase space for Example 5. b) The connection graph. c) The secondary inner stable division. d) The secondary outer stable division. e) Inner bridge class dynamics for Example 5.*

We can define a *1D stretching rate* for curves as $\lim_{n\to\infty}[\log(\ell_n)]/n$ where $\ell_n$ is the length of a curve after $n$ iterates. Similarly the *2D stretching rate* for surfaces is $\lim_{n\to\infty}[\log(a_n)]/n$ where $a_n$ is the area of a surface after $n$ iterates. The homotopic lobe dynamics of surfaces presented thus far provides a lower bound on the 2D stretching rate. As we shall see, the homotopy theory of curves presented below provides a lower bound on the 1D stretching rate. One would expect from dynamics that "mixes" in 3D that the 2D stretching rate would be strictly greater than the 1D stretching rate.

For maps with an axis of rotational symmetry, the symbolic dynamics of curves and the symbolic dynamics of surfaces are exactly the same, since all of the dynamics is captured in a cross-sectional plane, as in Example 2 (Sect. 2.2.) When this symmetry is broken, however, the symbolic dynamics of curves and surfaces may differ. After doing the hard work of deriving the symbolic dynamics of surfaces, we can simplify these results to obtain the symbolic dynamics of curves. We illustrate this for Examples 3 and 5.

**4.1. 1D homotopy analysis for Example 3.** Consider a non-self-intersecting curve that lies within a (two-dimensional) bridge, beginning and ending on the stable cap $T^S$, but otherwise not intersecting $T^S$. We call such a curve a 1D bridge. The homotopy class of a 1D bridge in 3D is uniquely determined from its homotopy class within the 2D bridge it lies. For example, the 2D bridge class $[\![A]\!]$ in Fig. 20 is represented by a topological disk. Any 1D bridge within this disk connects the disk boundary to itself, and is clearly contractible. Thus, the only 1D bridge class derived from $[\![A]\!]$ is the trivial bridge class. This is clearly true for any 1D bridge-class derived from a cap. More generally, any nontrivial 1D bridge must connect two distinct boundary classes of the 2D bridge within which it lies. We label a general 1D bridge class connecting a boundary class $X$ with a boundary class $Y$ by $[\![X,Y]\!]_1$. Possible inner 1D bridge classes obtained from Fig. 20 are: $[\![C,D]\!]_1$, $[\![B,C]\!]_1$, $[\![B,D]\!]_1$, $[\![A,E]\!]_1$, $[\![D,F]\!]_1$, $[\![C,F]\!]_1$, $[\![B,F]\!]_1$. There is a single nontrivial outer 1D bridge class: $[\![G,I]\!]_1$. Note that $[\![D,F]\!]_1$ is trivial because a curve connecting $D$ to $F$, without intersecting any of the bridges used to create the primary division, is contractible, since $D$ and $F$ lie within the same domain of the primary division in Fig. 19e. Similarly, $[\![C,D]\!]_1 = [\![C,F]\!]_1$ and $[\![B,D]\!]_1 = [\![B,F]\!]_1$. Furthermore, bridge classes $[\![C,D]\!]_1$ and $[\![A,E]\!]_1$ are equal to one another. This can be seen by the fact that both classes can be represented by the curve connecting $\mathbf{s}_0$ to $\mathbf{q}_1$ on the right side of Fig. 19a. In summary, there are four nontrivial 1D bridge classes

$$a = [\![C,D]\!]_1 = [\![A,E]\!]_1 = [\![C,F]\!]_1, \tag{18}$$

$$b = [\![B,D]\!]_1 = [\![B,F]\!]_1, \tag{19}$$

$$v_0 = [\![B,C]\!]_1, \tag{20}$$

$$u_0 = [\![G,I]\!]_1, \tag{21}$$

$$1 = [\![D,F]\!]_1. \tag{22}$$

As in Sect. 2.2, the symbols $a$, $b$, $v_0$, $u_0$ correspond to a particular orientation of curve, so that the ordering of boundary classes in the double brackets now matters, e.g. $a^{-1} = [\![D,C]\!]_1$.

To iterate the 1D classes forward, we consider the iterates of the 2D bridge classes in which they lie (Fig. 28). We map the boundary circles of the 1D classes forward, identify where they are in the corresponding graph on the left side of Fig. 28, and then connect them by the shortest path within the graph. For example, class $a$ connects $C$ to $D$, Eq. (18). Examining line 2 in Fig. 28, the iterate of $a$ connects $D$ to $E$ on the right side. Thus, the iterate of $a$ is $[\![D,C]\!]_1[\![I,G]\!]_1[\![A,E]\!]_1 = a^{-1}u_0^{-1}a$. The forward iterates of the other symbols can be analyzed in a similar way. In summary, we find

$$M(a) = a^{-1}u_0^{-1}a, \tag{23}$$
$$M(b) = a^{-1}u_0^{-1}a, \tag{24}$$
$$M(v_0) = 1, \tag{25}$$
$$M(u_0) = 1. \tag{26}$$

The inner class $v_0$ and the outer class $u_0$ are both inert, since they map to 1. So, we have the following transition matrix for the active 1D bridge classes

$$T = \begin{pmatrix} 2 & 2 \\ 0 & 0 \end{pmatrix}. \tag{27}$$

Note that symbol $b$ is transient, since nothing maps to it. Thus, Eq. (23) encodes the important (i.e. recurrent) 1D dynamics, which is simply the horseshoe dynamics from the right of Fig. 16a.

The log of the largest eigenvalue of Eq. (27) gives the stretching rate $h_{1\text{D}}$ of curves, as determined from the homotopy theory. This is a lower bound on the true stretching rate of curves for the 3D map. Obviously, we have

$$h_{1\text{D}} = \ln 2. \tag{28}$$

The topological entropy, determined previously in Eq. (13), is the stretching rate of 2D surfaces. In this example, the two stretching rates are equal. In this sense, the 3D nature of our topological model has not increased its complexity.

**4.2. 1D homotopy analysis for Example 5.** Following the same protocol as the previous section, we determine the possible 1D bridge classes from Fig. 36e. The inner classes are: $[\![C, F]\!]_1$, $[\![D, E]\!]_1$, $[\![A, F]\!]_1$, $[\![d, E]\!]_1$, and $[\![a, F]\!]_1$. Only two of these are unique, so we define

$$a = [\![C, F]\!]_1 = [\![A, F]\!]_1 = [\![a, F]\!]_1, \tag{29}$$
$$b = [\![D, E]\!]_1 = [\![d, E]\!]_1. \tag{30}$$

The forward iterates of these two classes, as determined from Fig. 37e, are

$$M(a) = b^{-1}u_0^{-1}a, \tag{31}$$
$$M(b) = b^{-1}v_0^{-1}a, \tag{32}$$

where $u_0 = [\![J, H]\!]_1$ and $v_0 = [\![J, I]\!]_1$ are outer bridge classes that are inert. The transition matrix for active 1D bridge classes is then

$$T = \begin{pmatrix} 1 & 1 \\ 1 & 1 \end{pmatrix}, \tag{33}$$

which yields the 1D stretching rate

$$h_{1\text{D}} = \ln 2. \tag{34}$$

Recall that we previously computed the 2D stretching rate, equal to the topological entropy, to be

$$h_{2D} = \ln 2.3593. \tag{35}$$

Therefore, Example 5 produces a 2D stretching rate that is strictly larger than the 1D stretching rate, meaning that there is no way in which this dynamics could be reduced to a 2D map.

**5. Conclusion.** We have demonstrated, through a series of examples, how homotopy theory can be used to extract a symbolic representation of three-dimensional dynamics from finite pieces of intersecting two-dimensional (codimension-one) stable and unstable manifolds. Future work will apply this technique to an explicit numerical mapping of three variables. Another future challenge will be to automate the extraction of the topological dynamics from the trellis. Such automation will be necessary to apply this technique to general and realistic problems, e.g. in chaotic 3D fluid flows, whose complexity quickly swamps one's ability to carry out by hand. Finally, an interesting question will be how these homotopic ideas extend to maps of dimension greater than three. Dimension four is particularly relevant for Hamiltonian systems.

**Acknowledgments.** This work has been gestating for several years and has benefitted from feedback and encouragement from numerous individuals, including Joshua Arenson, John Delos, Katherine Hess-Bellwald, Roy Goodman, Christof Jung, Hector Lomelí, Jim Meiss, Jay Mireles-James, and Jean-Luc Tiffeault.

**References.**